\documentclass[a4paper, 11pt]{article}
\usepackage{amsfonts}

\usepackage{amsmath,amsthm,amssymb,amscd,epsfig}
\usepackage{graphicx}
\usepackage{psfrag}

\newcommand {\R}{\mathbb{R}}

\newcommand {\rd}{\mathrm{d}}

\newcommand {\T}{\mathbb{T}}

\newtheorem{hypo}{Hypothesis}
\newtheorem{theorem}{Theorem}[section]
\newtheorem{proposition}[theorem]{Proposition}

\newtheorem{lemma}[theorem]{Lemma}
\newtheorem{remark}[theorem]{Remark}

\numberwithin{equation}{section}

\begin{document}

\title{Classical motion in force fields with short range correlations}
\author{
B. {\ Aguer}\thanks{Benedicte.Aguer@math.univ-lille1.f},
S. {\ De Bi\`{e}vre}\thanks{Stephan.De-Bievre@math.univ-lille1.fr},
P. {\ Lafitte}\thanks{Pauline.Lafitte@math.univ-lille1.fr} \\
Laboratoire Paul Painlev\'e, CNRS, UMR 8524 et UFR de Math\'ematiques \\
Universit\'e des Sciences et Technologies de Lille \\
F-59655 Villeneuve d'Ascq Cedex, France.\\
Equipe-Projet SIMPAF \\
Centre de Recherche INRIA Futurs \\
Parc Scientifique de la Haute Borne, 40, avenue Halley B.P. 70478 \\
F-59658 Villeneuve d'Ascq cedex, France.\\
and P.~E. {Parris}\thanks{parris@mst.edu} \\
Department of Physics\\
Missouri University of Science \& Technology,\\
Rolla, MO 65409, USA\\}
\date{\today}
\maketitle

\begin{abstract}
We study the long time motion of fast particles moving through time-dependent
random force fields with correlations that decay rapidly in space,
but not necessarily in time. The time dependence of the
averaged kinetic
energy $\left\langle p^{2}\left(t\right)\right\rangle /2$ and mean-squared displacement
$\left\langle q^{2}\left(t\right)\right\rangle $ is shown to exhibit a large
degree of universality; it depends only on whether the force is, or is not,
a gradient vector field. When it is, $\left\langle p^{2}\left(t\right)\right\rangle \sim
t^{2/5}$ independently of the details of the potential and of the space
dimension. Motion is then superballistic in one dimension,
with $\left\langle q^{2}\left(t\right)\right\rangle \sim t^{12/5}$, and ballistic in higher dimensions,
with $\left\langle q^{2}\left(t\right)\right\rangle \sim t^{2}$. These predictions are supported
by numerical results in one and two dimensions. For force fields not
obtained from a potential field, the power laws are
different: $\left\langle p^{2}\left(t\right)\right\rangle \sim t^{2/3}$
and $\left\langle q^{2}\left(t\right)\right\rangle \sim t^{8/3}$ in all dimensions $d\geq 1$.
\end{abstract}

\section{Introduction}
We study in this paper the motion
\begin{equation}
\ddot{q}\left(t\right)=F\left(q\left(t\right),t\right)  \label{eq:newton}
\end{equation}%
of fast particles in random force fields with correlations that
are short-range in space, but not necessarily in time.
We consider models of two different general classes. In the first,
upon which we focus most of our attention, the force is assumed to be of the form
\begin{equation}
F\left(q,t\right)=\sum_{N}f_{N}\left(\frac{q-q_{N}}{\ell },\frac{t}{\sigma }\right),
\label{eq:scatforce}
\end{equation}
where the $f_{N}$ are smooth functions of compact support in a ball of
radius $1/2$ centered at $0,$ with additional characteristics detailed in
Section \ref{s:prw}; $\ell, \sigma >0$ are a length and a time scale.
The $f_N$ model a random or periodic array of
identical, randomly-oriented scatterers,
centered at points $q_N$, that evolve periodically or
quasi-periodically in time. We assume
$
\inf_{N\neq M}\Vert q_{N}-q_{M}\Vert \geq \ell
$ so that the local forces $f_{N}\left(\frac{q-q_{N}}{\ell },\frac{t}{\sigma }\right)$ do
not overlap. As a result, the particle interacts with at most
one scatterer at a time, and otherwise travels freely between
collisions. The model therefore
describes an inelastic ``soft'' Lorentz gas, i.e., a
distribution of soft scatterers centered at the points $q_{N}$, off which
the particle bounces inelastically. We introduce randomness in the initial
data, and assume the system
to have finite horizon, so any trajectory of a free
particle intersects the support of $F$ at some future time $t$.

In the standard Lorentz gas, scattering is elastic, and scatterers are
identical hard unchanging obstacles centered at fixed points $q_{N}$, with a spatial
distribution chosen either randomly, periodically, or quasi-periodically.
Unlike the current model, energy in the Lorentz gas is conserved
and particle motion is diffusive \cite{bcs}. Another
diffusive model, related both to the Lorentz
gas and to those considered here, was
studied in \cite{spd, ldp}; there the scattering mechanism was
provided by a one-dimensional periodic array of oscillators representing
environmental degrees of freedom of the medium. The
Hamiltonian interaction of the particle with
the oscillator bath then provides, in addition to a random
force, an effective friction force that allows it to dissipate
the energy it gains, and to thus equilibrate with
its environment.

The force (\ref{eq:scatforce}) considered in the
present paper can be obtained from those of \cite{spd, ldp} by
switching off the friction component of the force provided by the
particle's back reaction with the medium.
The stochastic acceleration exerted on the particle by the random
force field leads, then, to an unbounded acceleration of the particle,
and it is of interest to compute the power laws associated, e.g.,
with the growth in time of the particle's average kinetic
energy $\left\langle p^2\left(t\right)\right\rangle /2$ and mean-squared displacement
$\left\langle q^2\left(t\right)\right\rangle$.

In the other class of models that we consider, the force $F\left(q,t\right)$ is
modeled as a space and time homogeneous random field satisfying
\begin{equation}
\left\langle F\left(q,t\right)\right\rangle =0,\qquad \left\langle F\left(q,t\right)F\left(q^{\prime },t^{\prime
}\right)\right\rangle
=\frac{\ell^2}{\sigma^4}C\left(\frac{q-q^{\prime }}{\ell },\frac{t-t^{\prime }}{\sigma }\right),
\label{eq:correlatedF}
\end{equation}
where $C$ is a matrix function of rapid decay in the spatial variable, but need not
decay in the time variable.  For these models, as with (\ref{eq:scatforce}),
we are interested in characterizing the asymptotic growth of
$\left\langle p^2\left(t\right)\right\rangle$ and $\left\langle q^2\left(t\right)\right\rangle$.

There has been a fair amount of work reported in the physics and mathematical
physics literature on problems of this type, partially motivated by questions in
plasma physics, astronomy, and solid state physics
(see for example \cite{st, sg, ve, bmv}).
Previous mathematically
rigorous work has mostly dealt with deriving, under suitable scalings,
Fokker-Planck equations for the particle
density (as in \cite{pv, gr}). Unfortunately, analyses of this type
do not directly give information about the asymptotic
behavior of the particles' kinetic energy or mean squared displacement.
The theoretical physics literature is mostly concerned with Gaussian random potentials
and contradictory
claims have been made regarding the power law growth
of $\left\langle p^2\left(t\right)\right\rangle$ and $\left\langle q^2\left(t\right)\right\rangle$. For
potential fields that are delta correlated in
time, but not in space, it is generally
agreed (see for example \cite{jk}) that
in the weak coupling limit $\left\langle p^2\left(t\right)\right\rangle\sim t$
and $\left\langle q^2\left(t\right)\right\rangle\sim t^3$, but there is some
controversy on what happens when the Gaussian potential field has temporal
correlations of nonzero and finite duration. For this case it is
argued in \cite{gff, lmf, r} that
in $d=1$, $\left\langle p^2\left(t\right)\right\rangle\sim t^{2/5}$ and
that $\left\langle q^2\left(t\right)\right\rangle\sim t^{12/5}$ (compatible
with numerical and theoretical results presented here).
In \cite{h92}, on the other hand, it is claimed that for $d=1$,
$\left\langle q^2\left(t\right)\right\rangle\sim t^3$,
as in the case when the random potentials are
delta correlated in time.
For $d>1$ it is found in \cite{gff} that
$\left\langle p^2\left(t\right)\right\rangle\sim t^{1/2}$, and
that $\left\langle q^2\left(t\right)\right\rangle\sim t^{9/4}$. In \cite{r}, the conclusions of
\cite{gff} for $d>1$ are contested and it is argued that for
Gaussian random potentials with fast decaying spatial
and temporal correlations $\left\langle p^2\left(t\right)\right\rangle\sim t^{2/5}$ in all dimensions,
and $\left\langle q^2\left(t\right)\right\rangle\sim t^{2}$ for $d>1$.

\begin{figure}[tbp]
\begin{center}
\includegraphics[width=12cm,keepaspectratio=true]{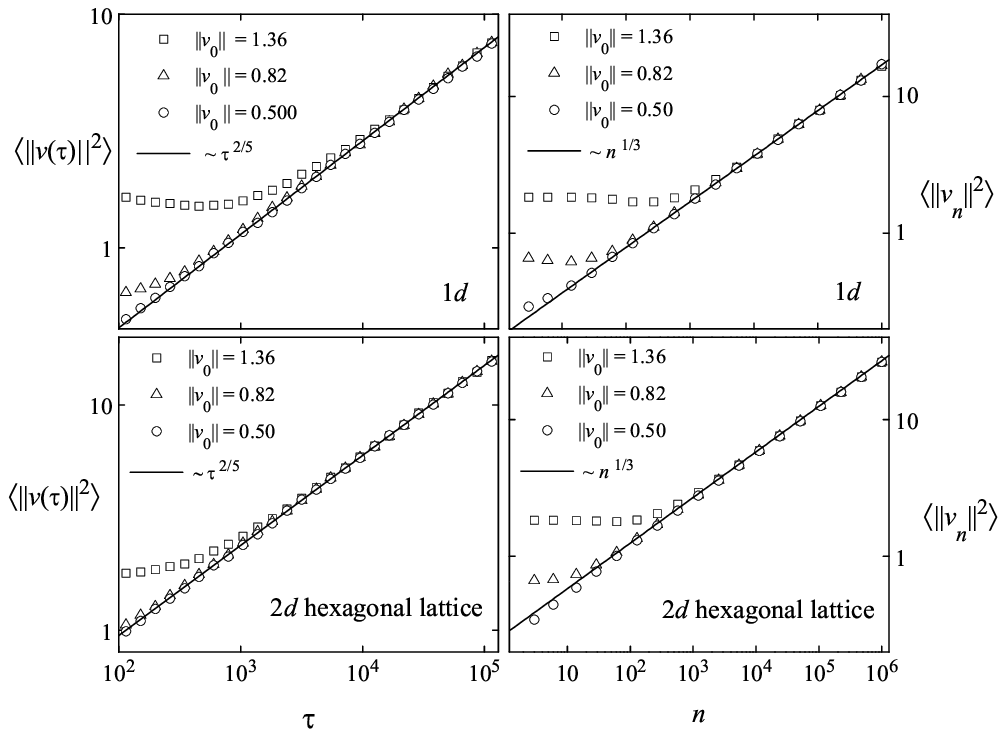} %
\end{center}
\caption{Numerically determined values of
$\left\langle v^2\left(\tau\right)\right\rangle$ and $\left\langle v_n^2\right\rangle$
in one dimension (top) and for a two-dimensional
hexagonal lattice (bottom), for the model described in Section~\ref{s:num}.
The different symbols correspond to different
initial conditions, as indicated.}
\label{fig:figurev2}
\end{figure}

\begin{figure}[tbp]
\begin{center}%
\includegraphics[height=10cm,keepaspectratio=true]{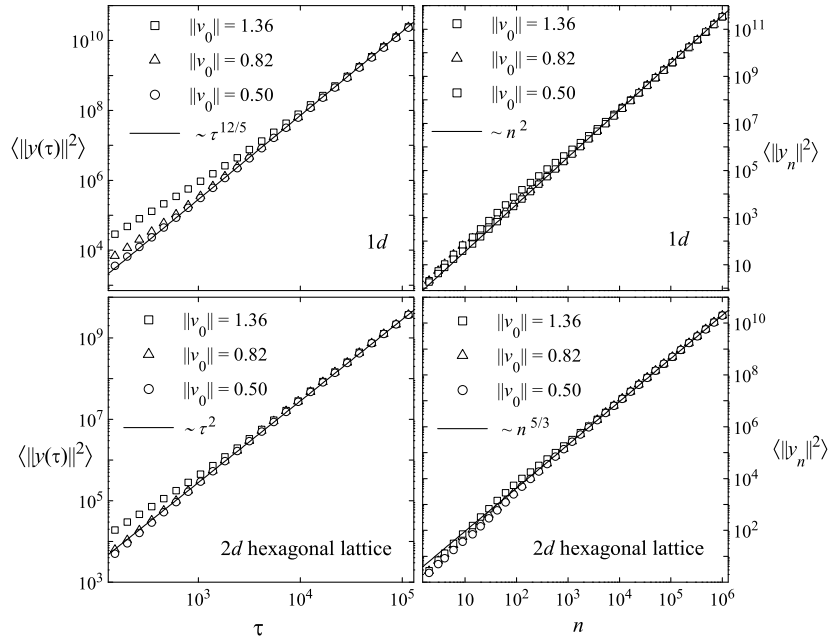}
\end{center}
\caption{Numerically determined values of
$\left\langle y^2\left(\tau\right)\right\rangle$ and $\left\langle y_n^2\right\rangle$, in one dimension (top), and for a two-dimensional
hexagonal lattice (bottom), for the model described in Section~\ref{s:num}.
In each plot, different symbols correspond to different
initial conditions, as indicated.}
\label{fig:figureq2}
\end{figure}

Although there is some numerical work \cite{lmf} that supports
the predictions for $d=1$ of \cite{gff, lmf, r}, to the best of our
knowledge no numerical simulations have been performed in
higher dimensions.
To help resolve the existing controversy on
this subject we present in this paper numerical results in
one and two dimensions on a particularly simple
(non-Gaussian) model whose random
force can be expressed as in (\ref{eq:scatforce}), and which allows for an efficient numerical
integration of the equations of motion out to very long times. Full details
of the numerical calculations are presented in Section~\ref{s:num},
but our essential results
for the case in which the force $F$ is derived from a potential field are
presented in Figures~\ref{fig:figurev2} and~\ref{fig:figureq2}, where we plot
the quantities $\left\langle v^2\right\rangle = \left\langle \left(p\sigma/\ell\right)^2\right\rangle$ and
$\left\langle y^2\right\rangle = \left\langle \left(q/\ell\right)^2\right\rangle$,
as functions both of the dimensionless time $\tau = t/\sigma$ and of
the collision number $n$, which labels
the number of scattering centers visited by the particle.

Our numerical results
indicate that in both one and two dimensions
\begin{equation}  \label{eq:asp2t}
\left\langle v^2\left(\tau\right)\right\rangle\sim \tau^{2/5},\qquad \left\langle v^2_{n}\right\rangle\sim n^{1/3},
\end{equation}
which is in agreement with \cite{gff, lmf, r}.
In one dimension the particle's mean-squared displacement
is superballistic, with
\begin{equation}  \label{eq:asq2t1}
\left\langle y^2\left(\tau\right)\right\rangle \sim \tau^{12/5},\qquad \left\langle y^2_{n}\right\rangle \sim n^{2}.
\end{equation}
In two dimensions, however, $\left\langle y^2\left(\tau\right)\right\rangle$ becomes ballistic, i.e.,
\begin{equation}  \label{eq:asq2td}
\left\langle y^2\left(\tau\right)\right\rangle \sim \tau^2, \qquad \left\langle y^2_{n}\right\rangle \sim n^{5/3}.
\end{equation}
This is different from what was predicted in \cite{gff} for Gaussian potentials, but in agreement
with predictions made for this case in \cite{r}.

To understand our numerical results in one and two dimensions, and to more firmly
establish what happens for the models of the type (\ref{eq:scatforce}) and (\ref{eq:correlatedF}) in higher
dimensions, we present in the
bulk of this paper a unified mathematical analysis that captures
the essential physics of the problem. It  provides in particular a
means for calculating the power law growth of the mean kinetic energy and
the mean-squared displacement associated with an ensemble of particles moving in
time-dependent random force fields of the types described above.

The analysis is based
on consideration of the typical trajectory of a particle moving in a
fluctuating force
field described by (\ref{eq:scatforce}) or (\ref{eq:correlatedF}), which can be viewed as a sequence
of isolated scattering events. We argue, in fact, that the motion is well
approximated by a coupled discrete-time random walk
for the particle's momentum and position. Each time step
corresponds to one collision of the particle with a single scatterer, or
to one traversal by the particle of a distance of the order of the
correlation length of the potential. Momentum increments are treated as
independent random events whose magnitude depends upon the particle
speed. Theoretical analysis of the resulting random walk reveals that
the high velocity behavior of the
momentum change of the particle during one such scattering event
completely determines the asymptotic properties of the motion.
As we show, this high energy
behavior is insensitive to the details of the force field,
notably to its statistical properties or to the precise geometry of
the scattering centers; the asymptotic behavior of the
motion is therefore quite universal, and in particular not a result
that arises only with Gaussian potential fields.

Indeed, for general force fields obtainable
as the gradient of a potential field,
we find (Theorem~\ref{propo_2})
that the energy change incurred by a particle of
velocity $v$ satisfies
$
{\Delta E}\sim \|v\|^{-1}.
$
This fact, combined with our analysis of the resulting random walk
in momentum and position space leads to an increase
of $\left\langle p^{2}\left(t\right)\right\rangle$ that is in all dimensions
of the form observed in Figure~\ref{fig:figurev2},
and as described by (\ref{eq:asp2t}). In one dimension,
$\left\langle q^{2}\left(t\right)\right\rangle$ is predicted by our analysis to grow in
time as in (\ref{eq:asq2t1}), and as observed
in the top left panel of Figure~\ref{fig:figureq2}. In all
higher dimensions it is predicted to
grow as observed in the bottom left panel in Figure~\ref{fig:figureq2},
and as described by (\ref{eq:asq2td}).  This slower growth
of $\left\langle q^{2}\left(t\right)\right\rangle$
in higher dimensions arises from the fact that the particle
can now turn while traveling, as its
velocity vector performs an orientational random walk resulting from
small random deflections.

Our analysis can also be applied to the case where $F(q,t)$ does not
derive from a potential field, a situation which has attracted some attention in the
mathematics literature. We find for a non-gradient force field that the
energy change
in a single scattering is considerably larger than
in the gradient case :
$
\Delta E\sim 1
$
(Theorem~\ref{propo_3}).
Consequently, we predict a larger rate of acceleration
$
\left\langle p^2\left(t\right) \right\rangle \sim t^{2/3}
$
(see (\ref{eq:diffvtnow})), that confirms rigorous results that have been obtained
for $d \geq 4$ in \cite{dk, kp79}
under suitable technical conditions on the
forces $f_N$ in (\ref{eq:scatforce}). Our analysis
then leads to the prediction that in all
dimensions particle motion in the presence of a non-gradient
random force field is superballistic with
$
\left\langle q^2\left(t\right) \right\rangle \sim t^{8/3}
$
(see (\ref{eq:diffqtnow})). The fundamental reason for the difference with the gradient field case
is that
the particle turns more slowly while traveling, because it accelerates more
quickly, and so is less easily deflected. The difference
between the two situations can be traced to the fact that
time-dependent gradient force fields produce
smaller changes in the particle's energy than
non-gradient force fields do. This is a remnant of the energy
conservation that is a characteristic feature of time-independent
gradient fields.

The rest of the paper is organized as follows. In Section~\ref{s:prw} we
introduce a random walk description of the motion of a particle moving
in a field of scatterers. General features of the walk that pertain to
both gradient and non-gradient
force fields are derived in Section~\ref{s:rwana}.
Section~\ref{s:gradient} is devoted to a derivation of
the above power laws for the case
of a gradient force field, and
Section~\ref{s:nongradient} analyzes the non-gradient
case. In Section~\ref{s:gfields}, we adapt our
analysis of Sections~\ref{s:rwana}-\ref{s:nongradient} to   random force fields
as described by (\ref{eq:correlatedF}), obtaining results
for the gradient and nongradient case identical to
those found, respectively, in Sections~\ref{s:gradient}
and \ref{s:nongradient}. Details of our numerical calculations,
the results of which are presented in figures distributed throughout the paper,
are given in Section~\ref{s:num}. Proofs of mathematical results used
for the analysis in Sections~\ref{s:rwana}-\ref{s:gfields}
comprise the Appendix.


\section{Particle in a field of scatters: a random walk description}
\label{s:prw}

 We first describe precise conditions on
the functions $f_N$ in (\ref{eq:scatforce}) under which we
work. We systematically use rescaled variables ($\ell>0, \sigma>0$)
\begin{equation*}
\tau=\frac{t}{\sigma}\in \R,
\quad y\left(\tau\right)=\frac{q\left(t\right)}{\ell}\in \R^d,
\quad v\left(\tau\right)=\dot y\left(\tau\right)=\frac{\sigma}{\ell}p\left(t\right),
\quad x_N = \frac{q_N}{\ell}
\end{equation*}
and suppose $f_N$ to be of the form
\begin{equation}  \label{eq:fN}
f_N\left(y,\tau\right)= \frac{\ell}{\sigma^2}c_N M_Ng\left(M_N^{-1}y,\omega \tau+\phi_N^0\right).
\end{equation}
The locations $x_N, N\in\mathbb{Z}^d$ of the scattering centers
can be chosen either randomly (with uniform density) or lying on
a regular lattice. The
coupling constants $c_N$ are independent random variables taking values in $%
[-1,1]$ and distributed according to a common probability measure $\nu$ not concentrated
on $0$. The $M_N$ are
rotations belonging to $\mathrm{SO}\left(d,\mathbb{R}\right)$ and are
also i.i.d., according to the
left-invariant Haar measure on $\mathrm{SO}\left(d,\mathbb{R}\right)$. Thus, the
scatterers are identical objects randomly oriented in space, all described
by the same function
$
g:\mathbb{R}^d\times \mathbb{T}^m\to\mathbb{R}^d
$
which is smooth and supported in the ball of radius $1/2$ in its first variable; $%
\mathbb{T}^m=\mathbb{R}^m/\mathbb{Z}^m$ is the $m$-torus and $\omega\in%
\mathbb{R}^m, \|\omega\|=1$. When $\omega$ has components that are
independent over the rationals, the force is quasi-periodic in
time, otherwise it is periodic.
The parameters $\phi_N^0\in\mathbb{T}^m$ are
i.i.d. random initial phases, uniformly distributed on the torus. We
write $\mathrm{d}\mu\left(M,\phi,c\right)$ for the above
described probability measure on $\mathrm{SO}\left(d,\mathbb{R}\right)\times\mathbb{T}^m%
\times[-1,1]$. The force may or may not derive from a potential. The
above class of models is sufficiently rich to allow for the description of
pulsing, vibrating, and rotating scattering centers; for an explicit example, see Section~\ref{s:num}.

In the rescaled variables, the equations of motion
(\ref{eq:newton})-(\ref{eq:scatforce}) become
\begin{equation}  \label{eq:rescnewton}
\ddot y\left(\tau\right)= G\left(y\left(\tau\right), \tau\right),\quad G\left(y,\tau\right)=\sum_N
c_NM_Ng\left(M_N^{-1}\left(y-x_N\right), \omega\tau+\phi_N^0\right).
\end{equation}
One should think of $g\left(y,\omega\tau+\phi\right)$ as the force produced by a soft,
time-dependent scatterer centered at the origin; $G$ then describes a
\emph{field} of \emph{identical} scatterers, randomly oriented, and
centered at the points $x_N$. We assume the system has a
finite horizon, so that the distance over
which a particle can freely travel is less than some fixed distance $L>0$,
uniformly in time and space and independently of the
direction in which it moves. Thus, with probability one, for
all $\left(y,v,\tau\right)\in\mathbb{R}^{2d}\times \mathbb{R}$ such
that $G\left(y,\tau\right)=0$,
\begin{equation*}
\sup\{\tau^{\prime}>0 \mid \forall 0\leq \tau^{\prime\prime}\leq
\tau^{\prime}, G\left(y+v\tau^{\prime\prime},\tau+\tau^{\prime\prime}\right)=0\}\leq
\frac{L}{\| v\|}.
\end{equation*}
\begin{figure}[tbp]
\begin{center}
\includegraphics[height=8cm]{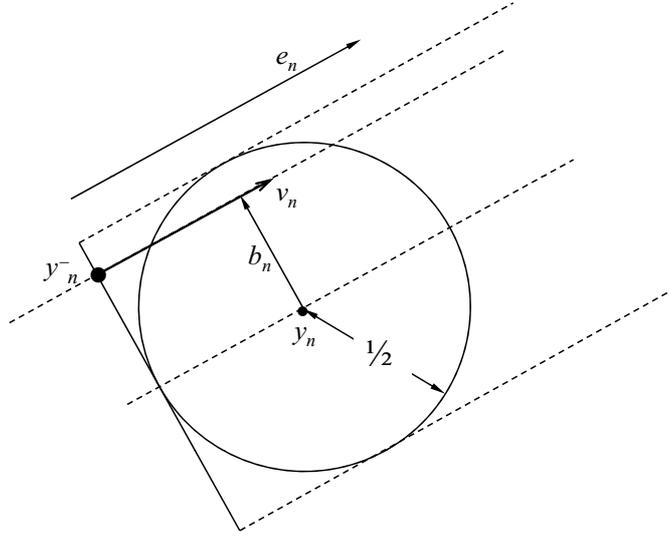}
\end{center}
\caption{A particle at time $\tau_n$ impinging with velocity $v_n$ and
impact parameter $b_n$ on the $n$th scatterer, centered at the
point $y_n$.}
\label{fig:collision}
\end{figure}
We consider a particle which at time $\tau_0=0$ is
close to a scatterer at $x_0=0$ and moving toward it with
initial velocity $v_0$ along an initial direction that, if
followed without deflection, would find the particle at
its closest approach to the force center
located at a point defined by the impact
parameter $b_0 \in \R^d$ (see Figure~\ref{fig:collision}).
After inelastically scattering from the center at $x_0$ the
particle moves freely with a new velocity $v_1$ until it
encounters a second scatterer, and
in this way it undergoes a random succession of scattering events.
The $n$th scattering event begins, by definition, at time $\tau_n$
when the particle arrives with incoming
velocity $v_n$ at the point (see Figure~\ref{fig:collision})
$$
y_n^- = y_{n} - \frac12 e_n + b_n,\qquad b_n\cdot e_n=0, \qquad\|b_n\|\leq \frac12
$$
near the scattering center at $y_n=x_{N_n}$,
where $e_n=v_n/\|v_n\|$, and the impact
parameter $b_n$ is a vector perpendicular to
the incoming velocity vector.
The $n$th scatterer itself is characterized
by its orientation $M_n:=M_{N_n}$, its
phase $\phi_n:=\omega\tau_n + \phi_{N_n}^0$ at the
time that the particle encounters it,
and the coupling strength $c_n:=c_{N_n}$.

The change in velocity experienced by a sufficiently fast particle
at the $n$th scattering center can be
written (Proposition \ref{prop:deltap})
\begin{equation}  \label{eq:vitrw}
v_{n+1}=v_n+R\left(v_n,b_n, M_n,\phi_n, c_n\right)
\end{equation}
where, for all $v\in\mathbb{R}^d$, $b\in\mathbb{R}^d$ with $v\cdot
b=0$, and $\left(M,\phi,c\right)\in \mathrm{SO}\left(d,\mathbb{R}\right)\times\mathbb{T}%
^m\times\mathbb{R}$,
\begin{equation}  \label{eq:momtransfer}
R\left(v, b, M,\phi,c\right)= c\int_{0}^{+\infty}\mathrm{d} \tau^{%
\prime}Mg\left(M^{-1}y\left(\tau^{\prime}\right), \omega\tau^{\prime}+\phi\right)
\end{equation}
in which $y\left(\tau\right)$ is the unique solution of
\begin{equation*}
\ddot y\left(\tau\right) = cMg\left(M^{-1}y\left(\tau\right),\omega\tau+\phi\right),
\quad\ y\left(0\right)=b-\frac12\frac{v}{\| v\|},\quad \ \dot
y\left(0\right)=v.
\end{equation*}
After leaving the influence of the $n$th scatterer, the particle
then travels a distance $\eta_n$ with velocity $v_{n+1}$ to
scatterer $n+1$, which it encounters after a
time $\Delta\tau_n = \eta_n/\|v_{n+1}\|$.

Based upon this description of the dynamics, and ignoring the role
of recollisions, we now argue that the motion of an ensemble of
particles moving in a force field described by (\ref{eq:scatforce})
is well approximated
by a coupled discrete-time random walk in momentum and position space.
Each step of the walk is associated with one scattering event,
where the variables ${M_n, \phi_n, c_n}$ that characterize the
scatterer, and the variables ${\eta_n, b_n}$
that characterize the approach of the particle onto the scatterer,
are drawn from distributions that characterize them
in the actual system of interest. Thus, starting from a given initial condition
$\left(y_0,v_0\right)$, we iteratively determine the velocity, the location, and the time
of the particle immediately before the $n$th scattering event through the relations:
\begin{equation}  \label{eq:finalrw}
\left.
\begin{array}{lll}
v_{n+1} & = & v_n+R\left(v_n, \kappa_n\right) \\
\tau_{n+1} & = & \tau_n+\frac{\eta_*}{\| v_{n+1}\|} \\
y_{n+1} & = & y_n+ \eta_* e_{n+1}%
\end{array}%
\right\}
\end{equation}
where $\kappa_n=\left(b_n,M_n,\phi_n,c_n\right)$. The parameters $\left(M_n,\phi_n,c_n\right)$ are
independently chosen from the distributions
already described (the distribution for $\phi_n$ being
the same as for $\phi_n^0$). Without the loss
of any essential physics, we have in (\ref{eq:finalrw}) replaced the
random variable $\eta_n$ at each time step
with the average distance $\eta_*=\left\langle\eta_n\right\rangle < L$ between
scattering events. The $b_n$ are
independently chosen at each step uniformly from the $d-1$ dimensional ball of
radius $1/2$ perpendicular to $v_n$.  To summarize, this random walk
describes a particle that moves freely over a distance $\eta_*$, then
meets, with random impact parameter, a randomly oriented scatterer
at a random moment of its (quasi-)periodic evolution. After
scattering, the process repeats itself. Our basic assumption, therefore,
is that this gives a good description of a typical trajectory
in the real system.

In what follows we write $\left\langle \cdot \right\rangle$ for
averages over all realizations of the random process $\kappa_n$.
In Sections~\ref{s:rwana}-\ref{s:gfields} we study the asymptotic
behavior of this random walk, under conditions expressed
in the following hypothesis:
\begin{hypo}
\label{hyp_Wtilde}
$g \in C^{3}\left(\mathbb{R}^d\times\mathbb{T}^m\right)$ is compactly
supported in the ball of radius $1/2$ centered at the origin in the $y$
variable. The function g and its partial derivatives up to order three are
all bounded, and we write
\begin{equation*}
0<g_{\mathrm{max}}:=\|g\|_{\infty}<+\infty.
\end{equation*}
If $g\left(y,\phi\right)=-\nabla_y W\left(y,\phi\right)$, we suppose $W\in C^4\left(\mathbb{R}^d\times
\mathbb{T}^m\right)$ also is supported in the ball of radius $1/2$ centered at the
origin in the $y$-variable. Moreover, $\left(\omega\cdot\nabla_\phi\right) W\not=0$
and, if $d=1$, we require that, for some $\phi\in\T^m$,
\begin{equation}  \label{eq:intnotzero}
\int_{\mathbb{R}} \mathrm{d} y\ \left(\omega\cdot\nabla_\phi\right) W\left(y,\phi\right)\not=0.
\end{equation}

\end{hypo}

\section{Analysis of the random walk: general considerations}
\label{s:rwana}

We now turn to the analysis of the large $n$ behavior of
the first equation of (\ref{eq:finalrw})
\begin{equation}  \label{eq:velrw}
v_{n+1}=v_n+R\left(v_n, \kappa_n\right)
\end{equation}
which is independent of the others. We assume that the particles are
fast, meaning $\|v_0\|^2>> cg_{\mathrm{max}}$ (Lemma \ref{lem:estim_ts}). For
that purpose we need to understand
the high momentum behavior of the momentum transfer $R\left(v_n, \kappa_n\right)$, as
well as of the energy transfer
\begin{equation}  \label{eq:deltaendef}
\Delta E\left(v_n, \kappa_n\right)=\frac12\left(\left(v_n+R\left(v_n, \kappa_n\right)\right)^2-v_n^2\right).
\end{equation}
First order perturbation theory allows one to write (see
Proposition \ref{prop:deltap} for details)
\begin{eqnarray*}
R\left(v_n, \kappa_n\right)&=&\frac{c_n}{\| v_n\|}\int_{-\infty}^{+\infty} \mathrm{d} \lambda \ M_ng\left(M_n^{-1}\left(b_n+\left(\lambda-\frac12\right) e_n\right), \frac{\omega\lambda}{\| v_n\|}%
+\phi_n\right)\\
&\ & \qquad\qquad\qquad\qquad\qquad\qquad\qquad\qquad\qquad\qquad  + \mathrm{O}\left(\| v_n\|^{-3}\right).
\end{eqnarray*}
More generally, if $g$ is sufficiently smooth, one can write, for $K\in\mathbb{N}$, and $\left(v,
\kappa\right)\in\mathbb{R}^{2d}\times\mathrm{SO}\left(d,\mathbb{R}\right)\times\mathbb{T}%
^m\times\mathbb{R}$, $b\cdot v=0$,
\begin{equation}  \label{eq:Rexpansion}
R\left(v, \kappa\right)=\sum_{k=1}^K \frac{\alpha^{\left(k\right)}\left(e, \kappa\right)}{\|v\|^k} +\mathrm{O}%
\left(\|v\|^{-K-1}\right),\quad e=\frac{v}{\|v\|}.
\end{equation}
Note that
\begin{equation}  \label{eq:alpha1}
\alpha^{\left(1\right)}\left(e, \kappa\right)={c}\int_{-\infty}^{+\infty} \mathrm{d} \lambda \
Mg\left(M^{-1}\left(b+\left(\lambda-\frac12\right) e\right), \phi\right)
\end{equation}
and
\begin{equation}  \label{eq:alpha2}
\alpha^{\left(2\right)}\left(e, \kappa\right)={c}\int_{-\infty}^{+\infty} \mathrm{d} \lambda \
\lambda\partial_\tau Mg\left(M^{-1}\left(b+\left(\lambda-\frac12\right) e\right), \phi\right),
\end{equation}
in which we have introduced the suggestive notation
\begin{equation}  \label{eq:partialtau}
\partial_\tau := \omega\cdot\nabla_\phi.
\end{equation}
Hence
\begin{equation}  \label{eq:deltaEexpansion}
\Delta E\left(v, \kappa\right)=\sum_{\ell=0}^L \frac{\beta^{\left(\ell\right)}\left(e, \kappa\right)}{%
\|v\|^{\ell}}+\mathrm{O}\left(\|v\|^{-L-1}\right),
\end{equation}
where
\begin{equation}
\left.
\begin{array}{lll}
\label{eq:betaell} \beta^{\left(0\right)} & = & e\cdot \alpha^{\left(1\right)} \\
\beta^{\left(1\right)} & = & e\cdot \alpha^{\left(2\right)} \\
\beta^{\left(2\right)} & = & \left(\frac12\alpha^{\left(1\right)}\cdot\alpha^{\left(1\right)}+e\cdot%
\alpha^{\left(3\right)}\right) \\
\beta^{\left(3\right)} & = & \left(\alpha^{\left(1\right)}\cdot\alpha^{\left(2\right)}+e\cdot\alpha^{\left(4\right)}%
\right) \\
\beta^{\left(4\right)} & = & \left(\frac12\alpha^{\left(2\right)}\cdot\alpha^{\left(2\right)}+
\alpha^{\left(1\right)}\cdot\alpha^{\left(3\right)}+e\cdot\alpha^{\left(5\right)}\right).%
\end{array}
\right\}
\end{equation}

It is easy to see that expansion (\ref{eq:Rexpansion}) has rather different
features when $g$ is a gradient vector field than when it is not.
Indeed, when $g=-\nabla W$,
the first order term in the momentum transfer (\ref{eq:Rexpansion}) is
perpendicular to the incoming momentum $v$, so that
\begin{equation}  \label{eq:betazerozero}
\beta^{\left(0\right)}\left(e, \kappa\right)=e\cdot \alpha^{\left(1\right)}\left(e, \kappa\right)=0.
\end{equation}
As a result $\Delta E\sim \|v\|^{-1}$ in that case. Moreover, one then has
\begin{equation}  \label{eq:beta1}
\beta^{\left(1\right)}\left(e, \kappa\right)=c\int_{-\infty}^{+\infty} \mathrm{d}\lambda\ \partial_\tau
W\left(M^{-1}\left(b+\lambda e, \phi\right)\right).
\end{equation}
On the other hand, when $g$ is not a gradient vector field, $\beta^{\left(0\right)}$
does not vanish and, as a consequence, $\Delta E\sim 1$. This is the source
of the different asymptotics for $\left\langle v_n^2\right\rangle$ and $\left\langle
y_n^2\right\rangle$ in those two cases, as we will see below.

For later purposes, starting from
(\ref{eq:velrw})-(\ref{eq:Rexpansion}), a simple
computation yields
\begin{equation}  \label{eq:enplus1}
e_{n+1}=\left(1-\frac{\Delta E_n}{\|v_n\|^2}\right)\left[e_n + \frac{R_n}{%
\|v_n\|}\right]+ \mathrm{O}\left(\frac{\left(\Delta E_n\right)^2}{\|v_n\|^4}\right)
=e_n+\delta_n,
\end{equation}
where $R_n=R\left(v_n,\kappa_n\right)$, and $\Delta E_n=\Delta E\left(v_n,\kappa_n\right)$. Hence,
from (\ref{eq:betaell}),
\begin{eqnarray*}
\delta_n&=&\left(\alpha^{\left(1\right)}_n -\left(\alpha^{\left(1\right)}_n\cdot e_n\right)e_n\right)\frac{1}{\|v_n\|^2}
+\left(\alpha^{\left(2\right)}_n-\left(%
\alpha^{\left(2\right)}_n\cdot e_n\right)e_n\right)\frac1{\|v_n\|^3}  \notag \\
&\ &\qquad\quad +\left(\alpha^{\left(3\right)}_n-\left(\alpha^{\left(3\right)}_n\cdot
e_n\right)e_n\right)\frac1{\|v_n\|^4}
-\frac12\left(\alpha_n^{\left(1\right)}\cdot\alpha_n^{\left(1\right)}\right)\frac{e_n}{\|v_n\|^4} \\
&\ &\qquad\qquad\qquad\qquad\qquad\qquad + \mathrm{O}\left(\| v_n\|^{-5}\right)
 + \mathrm{O}\left(\frac{\left(\Delta E_n\right)^2}{\|v_n\|^4}\right) \notag \\
&=&\delta_n^{\left(4\right)}+\mathrm{O}\left(\| v_n\|^{-5}\right)
+ \mathrm{O}\left(\frac{\left(\Delta E_n\right)^2}{\|v_n\|^4}\right).
\end{eqnarray*}
Here $ \alpha^{\left(k\right)}_n=\alpha^{\left(k\right)}\left(e_n, \kappa_n\right). $ We can write
$\delta_n=\delta_n^\perp +\mu_ne_n, \ \delta_n^\perp\cdot e_n=0, $ with
(since $\|e_{n+1}\|=1=\|e_n\|$)
\begin{eqnarray*}
\mu_n&=&-1+\sqrt{1-\delta_n^\perp\cdot\delta_n^\perp}\leq 0 \\
&=&-\frac12\left(\alpha_n^{\left(1\right)}\cdot\alpha_n^{\left(1\right)}\right)\frac1{\|v_n\|^4}+%
\mathrm{O}\left(\| v_n\|^{-5}\right)
+ \mathrm{O}\left(\frac{\left(\Delta E_n\right)^2}{\|v_n\|^4}\right).
\end{eqnarray*}

For a
function $f$ depending on $v$ and $\kappa=\left(b,M,\phi,c\right)$, $b\cdot v=0, \|
b\|\leq 1/2$ we shall denote the average over parameters associated with
a single scattering event as
\begin{equation}  \label{eq:overlinedef}
\overline{f\left(v\right)}=\int \frac{\mathrm{d} b}{C_d}\int \mathrm{d}\mu\left(M,\phi,c\right)
f\left(v,b,M,\phi,c\right),
\end{equation}
where $C_d$ is the volume of the ball of radius $1/2$ in $\mathbb{R}^{d-1}$.

\section{Analysis of the random walk: gradient fields}
\label{s:gradient}

In this section we consider the more interesting case where $g=-\nabla_y W$.
The following theorem, the proof of which appears in the Appendix,
 will be essential to our results.
\begin{theorem}
\label{propo_2} Suppose Hypothesis \ref{hyp_Wtilde} holds and that $%
g=-\nabla_y W$.\newline
(i) For all unit vectors $e\in\mathbb{R}^d$,
\begin{equation}  \label{eq:alphaaverage}
\overline{\alpha^{\left(1\right)}\left(e\right)}=0=\overline{\alpha^{\left(2\right)}\left(e\right)}.
\end{equation}
Moreover, for all $v\in\mathbb{R}^d$
\begin{equation}  \label{eq:energyaverage}
\overline{\Delta E\left(v\right)}=\frac{B}{\|v\|^4} +\mathrm{O}\left(\|v\|^{-5}\right),\quad
\overline{\left(\Delta E\left(v\right)\right)^2}=\frac{D^2}{\|v\|^2} +\mathrm{O}\left(\|v\|^{-3}\right),
\end{equation}
where
\begin{equation}  \label{eq:bdlink}
B=\frac{d-3}{2}D^2
\end{equation}
with
\begin{equation}  \label{eq:d2formula}
D^2=\frac{\overline{c^2}}{C_d}\int_{\T^m}{\mathrm{d} \phi}
\int_{\mathbb{R}^{2d}} \mathrm{d} y_0 \mathrm{d} y_0^{\prime}\parallel
y_0-y_0^{\prime}\parallel^{1-d}\partial_\tau W\left(y_0,\phi\right)\partial_\tau
W\left(y_0^{\prime},\phi\right)>0.
\end{equation}
In particular, for all unit vectors $e\in\mathbb{R}^d$ and for $\ell=1,2,3$,
\begin{equation}  \label{eq:betaaverage}
\overline{\beta^{\left(\ell\right)}\left(e\right)}=0,\quad B=\overline{\beta^{\left(4\right)}\left(e\right)} \ \mathrm{%
and}\ D^2=\overline{ \left(\beta^{\left(1\right)}\left(e\right)\right)^2}>0.
\end{equation}
(ii) Let $v_n$ be the random process defined by (\ref{eq:velrw}) and $%
e_n=v_n/\|v_n\|$. Let, for $\ell\in\mathbb{N}$,
$
\beta_n^{\left(\ell\right)}=\beta^{\left(\ell\right)}\left(e_n, \kappa_n\right).
$
Then one has, for all $n\not=n^{\prime}\in \mathbb{N}$,
for all $0\leq
\ell\leq\ell^{\prime}\leq 3$,
\begin{equation}
\left.
\begin{array}{l}
\label{eq:betaellindep} \left\langle\beta^{\left(4\right)}_n\right\rangle-B=0=\left\langle\beta_n^{\left(%
\ell\right)}\right\rangle     \\
\left\langle
\beta_n^{\left(\ell\right)}\beta_{n^{\prime}}^{\left(\ell^{\prime}\right)}\right\rangle=0=\left\langle
\beta_n^{\left(\ell\right)}\left(\beta_{n^{\prime}}^{\left(4\right)}-B\right)\right\rangle=\left\langle\left(%
\beta_{n}^{\left(4\right)}-B\right)\left(\beta_{n^{\prime}}^{\left(4\right)}-B\right)\right\rangle.
\end{array}
\right\}
\end{equation}
Moreover, $\left\langle \left(\beta^{\left(4\right)}_n\right)^2\right\rangle$ and $\left\langle\beta^{\left(\ell\right)}_n%
\beta^{\left(4\right)}_n\right\rangle$ are independent of $n$.
\end{theorem}

\begin{remark}
\label{remarkpropo_2} (i) Note that part (i) of the Theorem does not
involve the random walk (\ref{eq:velrw}). It is a statement about the
functions $\alpha^{\left(\ell\right)}\left(e,\kappa\right),\beta^{\left(\ell\right)}\left(e,\kappa\right)$, viewed as random variables in $%
\kappa$. \newline
(ii) The strict positivity of $D^2$ is equivalent to the
requirement that $\beta^{\left(1\right)}$ does not vanish identically. This follows
from Hypothesis~\ref{hyp_Wtilde}, and notably from the nonvanishing of
the time derivative of the potential. This is as expected, since in a
time-independent potential, energy is conserved to all orders, so certainly
$\beta^{\left(1\right)}=0$. In one dimension, the extra assumption (\ref{eq:intnotzero})
is needed to ensure $\beta^{\left(1\right)}\not=0$: indeed,
when $d=1$, $\beta^{\left(1\right)}=0$
as soon as the potential has a vanishing spatial average. In that case, some
lower order term $\beta^{\left(\ell\right)}$ will not vanish and, as will be clear
from the discussion which follows, this would alter the
power laws of the stochastic acceleration. Such situations, which are
easily treated using the methods of this paper, will not be considered further.
\end{remark}

We first establish the asymptotic behavior of $\left\langle\|v_n\|^2\right\rangle$,
where $v_n$ is the stochastic process defined by (\ref{eq:velrw}).
We start from the expansion (\ref{eq:deltaEexpansion}) which
yields, respectively
\begin{eqnarray*}
\frac{\|v_{n+1}\|^2}{\|v_n\|^2}&=&1+\sum_{i=1}^4\frac{2\beta_n^{\left(i\right)}}{%
\|v_n\|^{i+2}} + \mathrm{O}\left(\|v_n\|^{-7}\right) \\
\frac{\|v_{n+1}\|}{\|v_n\|}&=&1+\sum_{i=1}^3\frac{\beta_n^{\left(i\right)}}{%
\|v_n\|^{i+2}} + \mathrm{O}\left(\|v_n\|^{-6}\right) \\
\|v_{n+1}\|-\|v_n\|&=&\sum_{i=1}^3\frac{\beta_n^{\left(i\right)}}{\|v_n\|^{i+1}}+\frac{%
\beta_n^{\left(4\right)}-\frac12\left(\beta_n^{\left(1\right)}\right)^2}{\|v_n\|^5}+ \mathrm{O}\left(\|v_n\|^{-6}\right)
\end{eqnarray*}
and consequently
\begin{eqnarray}  \label{eq:v3rw}
\Delta \|v_n\|^3&=&\|v_n\|^2\Delta \|v_n\|\left[1+\frac{\|v_{n+1}\|}{\|v_n\|}%
+\frac{\|v_{n+1}\|^2}{\|v_n\|^2}\right]  \notag \\
&=&\sum_{i=1}^3\frac{3\beta_n^{\left(i\right)}}{\|v_n\|^{i-1}}+\frac{%
3\left(\beta_n^{\left(4\right)}+\frac12\left(\beta_n^{\left(1\right)}\right)^2\right)}{\|v_n\|^3}+ \mathrm{O}%
\left(\|v_n\|^{-4}\right)  \notag \\
&=&3\beta_n^{\left(1\right)} + \frac{3\left(\beta_n^{\left(4\right)}+\frac12\left(\beta_n^{\left(1\right)}\right)^2%
\right)}{\|v_n\|^3}+\mathrm{O}_0\left(\|v_n\|^{-1}\right)+\mathrm{O}\left(\|v_n\|^{-4}\right).
\notag \\
\end{eqnarray}
Here the notation $\mathrm{O}_0\left(\|v_n\|^{-1}\right)$ means the term
is $\mathrm{O}\left(\|v_n\|^{-1}\right)$ and of zero average. Introducing
\begin{equation}  \label{eq:normalize}
\xi_n= \frac{\|v_n\|^3}{3D},\ \epsilon_n=\frac{\beta_n^{\left(1\right)}}{D}\
\mathrm{and}\ \gamma=\frac13\left(\frac{B}{D^2}+\frac12\right)=\frac16\left(d-2\right)\geq
-\frac16,
\end{equation}
we drop the error term in (\ref{eq:v3rw}) to obtain the one-dimensional
random walk
\begin{equation}  \label{eq:xirw}
\Delta \xi_{n}=\epsilon_n+\frac{\gamma}{\xi_n}\quad\mathrm{with}\quad\left\langle
\epsilon_n\right\rangle=0,\qquad \left\langle \epsilon_n^2\right\rangle=1
\end{equation}
in the variable $\xi_n$. Here the first term on the right hand side is
the dominant term of zero
average in (\ref{eq:v3rw}), whereas the second term is a systematic drift
term, and is its dominant term of non-zero average (when $\gamma\not=0$).

\begin{figure}[tbp]
\begin{center}
\includegraphics[width=9cm, keepaspectratio]{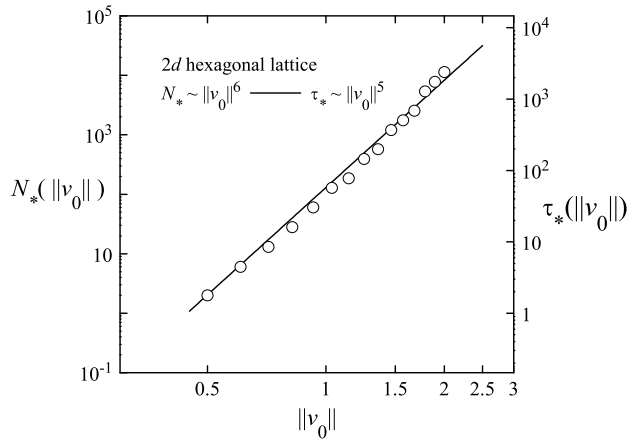}
\end{center}
\caption{The quantities $N_*\left(\|v_0\|\right)$ (on the left vertical axis), and
$\tau_*\left(\|v_0\|\right)$ (on the right vertical axis)
as a function of $\|v_0\|$ for a
particle moving in a hexagonal lattice $\left(d=2\right)$ as described in Section~\ref{s:num}.}
\label{fig:regimechange}
\end{figure}

From this simple random walk we can easily deduce the short time
behavior of the dynamics.
Suppose
$\xi_0>>|\gamma|$.
Then,
\begin{equation*}
\xi_n=\xi_0+n\frac{\gamma}{\xi_0} +\sum_{k=0}^{n-1}\epsilon_k,
\end{equation*}
where this approximation remains valid as long as
$
|\xi_n-\xi_0|<<\xi_0.
$
A short calculation shows this is guaranteed provided
\footnote{From this point onward we use the
notation $f\left(x\right)\sim g\left(x\right)$ to mean that there exist $0<c\leq
C<+\infty$ so that $cf\left(x\right)\leq g\left(x\right)\leq Cf\left(x\right)$.}
\begin{equation}  \label{eq:nstarxi}
n<<N_*\left(\xi_0\right) \sim \xi_0^2 \sim\ \|v_0\|^6.
\end{equation}
This last relation gives an estimate of the number of collisions needed
before the asymptotic long time behavior, as derived below, sets in.
This dependence on the initial speed can be seen in the numerical results for the model
described in Section~\ref{s:num}, as shown in Figure~\ref{fig:regimechange}.
We now turn to the asymptotic behavior of $\xi_n$ $n>>N_*\left(\xi_0\right)$.
We will show that, for $d\geq 2$, and for $k>-3$,
\begin{equation}  \label{eq:vnbehaviour}
\left\langle\|v_n\|^k\right\rangle\sim n^{\frac{k}{6}}.
\end{equation}

\begin{figure}[h]
\begin{center}
\includegraphics[height=9cm,keepaspectratio=true]{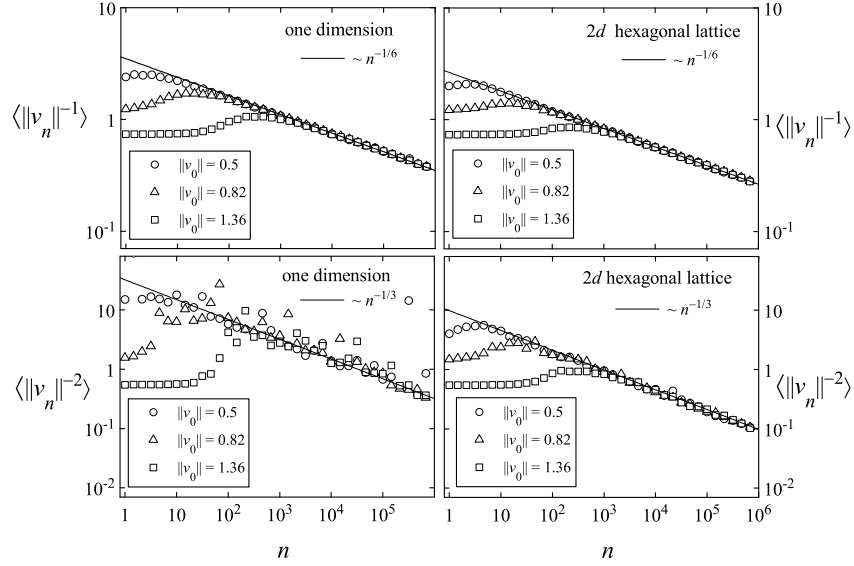}
\end{center}
\caption{Numerical results showing the asymptotic behavior of
$\left\langle\|v_n\|^k\right\rangle$, for the model described in Section~\ref{s:num}, in
one and two dimensions, with initial conditions and $k$ values as indicated.}
\label{fig:vnbehaviour}
\end{figure}

Note that this is indeed the behavior observed numerically for the full
dynamics of the numerical models described in Section~\ref{s:num}, as illustrated in
Figure~\ref{fig:figurev2} for $k=2$, and which we present in
Figure~\ref{fig:vnbehaviour} for $k=-1$ and $-2$.

From a theoretical point of view, the result (\ref{eq:vnbehaviour}) is obvious
for $d=2$, since
then $\gamma=0$ and (\ref{eq:xirw}) then just describes a simple
random walk on the half line.
More generally, looking at (\ref{eq:xirw}), because $\gamma\geq 0$
for $d\geq 2$, one certainly expects $\left\langle\xi_n\right\rangle\to+\infty$, as
a result of the
combined drift-diffusion implied by (\ref{eq:xirw}). In $d=1$, $\gamma<0$
and the second term then acts as a friction term. We will nevertheless show
that for all $\gamma > -1/2$ the friction is too small to alter
the asymptotic behavior of $\left\langle \xi_n\right\rangle$. To this end we note that
\begin{equation*}
\Delta \xi_{n}^2=\left(2\xi_n+\Delta\xi_n\right)\left(\Delta\xi_n\right).
\end{equation*}
Again keeping only the dominant terms yields
\begin{equation} \label{eq:xi2rw}
\Delta\xi_n^2=2\xi_n\epsilon_n + 2\gamma + 1.
\end{equation}

For $\gamma > -1/2$, we will
now show that after rescaling the process $\xi_n^2$ by $n$, it has a
well-defined limit, which is a squared Bessel process
of dimension $\delta=2\gamma+1$. This will
establish (\ref{eq:vnbehaviour}) for those values of $\gamma$ and for
all $k>{-3}$. To see this, define, for $s\geq 0$, $n\in\mathbb{N}$,
and $0\leq \sigma\leq s$,
\begin{equation}  \label{eq:bessel1}
Y_\sigma^{\left(n\right)} =\frac{s}{n}\xi_k^2,\quad\mathrm{if}
\quad \sigma_k=
k\frac{s}{n}%
\leq \sigma<\left(k+1\right)\frac{s}{n}=\sigma_{k+1}.
\end{equation}
Multiplying (\ref{eq:xi2rw}) with $s/n$ one finds that
\begin{equation*}
Y^{\left(n\right)}_s=Y_0^{\left(n\right)}
+ 2\sum_{k=0}^{n-1}\sqrt{Y_{\sigma_k}^{\left(n\right)}}\Delta B_{\sigma_k}^{\left(n\right)}
+ \left(2\gamma +1\right)s,
\end{equation*}
where
\begin{equation*}
B^{\left(n\right)}_{\sigma_k}=\sqrt{\frac{s}{n}}\sum_{\ell=0}^{k-1} \epsilon_{\ell}.
\end{equation*}
Taking the limit $n\to+\infty$ and writing $Y_s=\lim_{n\to+\infty}%
Y_s^{\left(n\right)}$, one finds
\begin{equation*}
Y_s=Y_0 + 2\int_0^s \sqrt{Y_s} \mathrm{d} B_s + \left(2\gamma+1\right)s,
\end{equation*}
where $B_s$ is a one-dimensional Brownian motion since the $\epsilon_n$ are
i.i.d. In other words, the limiting process $Y_s$ satisfies the stochastic
differential equation
\begin{equation}  \label{eq:bessel}
\mathrm{d} Y_s = 2\sqrt{Y_s}\mathrm{d} B_s + \left(2\gamma+1\right)\mathrm{d}s,
\end{equation}
of the squared Bessel process of dimension $\delta=2\gamma+1$ (see \cite{ry},
Chapter~11), and is therefore a squared Bessel process.

\begin{figure}[tbp]
\begin{center}
\includegraphics[width=12cm, keepaspectratio]{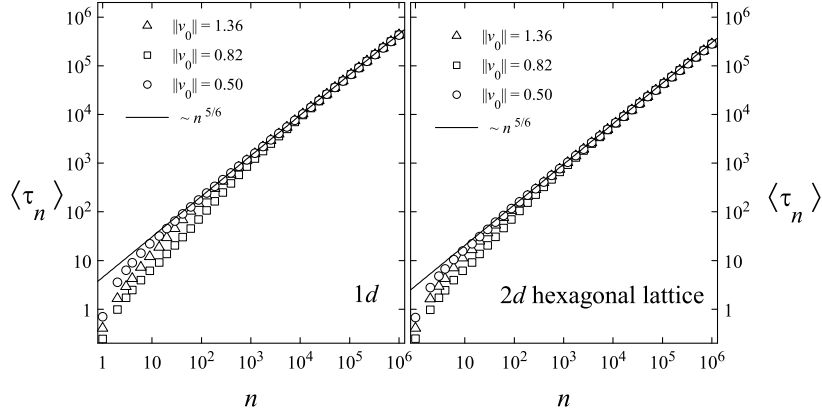}
\end{center}
\caption{Behavior of the collision time $\left\langle\tau_n\right\rangle$ for the model of
Section~\ref{s:num}, both in one and two dimensions, with initial speeds
as indicated.}
\label{fig:taun}
\end{figure}

Thus, since $\xi_n^2/n$ converges, we can approximate its distribution by
that of $Y_1$ and conclude that, for all $\ell>-1$,
$
\left\langle \xi_n^\ell\right\rangle \sim n^{\frac{\ell}{2}},
$
which is (\ref{eq:vnbehaviour}).
Equation (\ref{eq:vnbehaviour}) in particular yields,
via (\ref{eq:finalrw}), (see Figure~\ref{fig:taun})
\begin{equation}  \label{eq:taun}
\left\langle \tau_n\right\rangle \sim n^{5/6}
\end{equation}
and, finally
\begin{equation}  \label{eq:diffvtw}
\left\langle v^2\left(\tau\right)\right\rangle \sim \tau^{2/5},\quad \tau>> \tau_*\left(\|v_0\|\right):=
\frac{N_*\left(\xi_0\right)}{\|v_0\|}\sim \|v_0\|^5.
\end{equation}

We note that the asymptotic behavior does not depend on the initial speed $%
\|v_0\|$; the time scale $\tau_*\left(\|v_0\|\right)=\tau_{N_*\left(\|v_0\|\right)}$ on
which it sets in, on the other
hand, is predicted by this analysis to grow quickly,
as $\|v_0\|^5$, a result that is verified in the $2d$
numerical results presented in Figure~\ref{fig:regimechange}, which shows
the number of collisions
$N_*$ and the mean time $\tau_*$ before the asymptotic regime
is observed, as a function of $\|v_0\|$.

For $d=1$, the same power law was found for (Gaussian) random fields in \cite{gff}, \cite{lmf}, and
 \cite{awmn}, using very
different methods. For $d>1$, the only studies we are aware of
are \cite{gff} and \cite{r}, who deal with Gaussian random fields and who respectively
find $\left\langle v^2\left(\tau\right)\right\rangle \sim \tau^{1/2}$, which disagrees
with (\ref{eq:diffvtw}) and $\left\langle v^2\left(\tau\right)\right\rangle \sim \tau^{2/5}$, which agrees with it. As mentioned in the
Introduction, we have corroborated our predictions (\ref{eq:vnbehaviour})
and (\ref{eq:diffvtw}), including the
onset of the asymptotic regime
at $\tau_*\left(\|v_0\|\right)$, with numerical calculations in $1d$ and $2d$,
the results of which are presented in Figure~\ref{fig:figurev2}
and described more fully in Section~\ref{s:num}.

Before turning back to an analysis of the full random
walk (\ref{eq:finalrw}) in
order to determine the asymptotic behavior of the mean squared
displacement $\left\langle y^2\left(\tau\right)\right\rangle$, we now first briefly
discuss the validity of
the assumption we implicitly made in passing from (\ref{eq:v3rw}) to (\ref%
{eq:xirw}), namely that the lower order terms of (\ref{eq:v3rw}) won't alter
the behavior of $\left\langle v^2\left(\tau\right)\right\rangle$ that we obtained by
ignoring them.
To get an estimate of the error made neglecting these terms, we will
evaluate them along a typical trajectory of the random
walk (\ref{eq:velrw}), along which we
showed $\|v_n\|\sim n^{1/6}$, and will thereby demonstrate
that the contribution of each of the neglected terms to $\xi_n$ is smaller
than $n^{1/2}$, the contribution of the two dominant terms retained
above. Note first that for $i=2,3$
\begin{equation*}
\left\langle\sum_{k=1}^n \frac{\beta^{\left(i\right)}_k}{k^{\left(i-1\right)/6}}\right\rangle=0,
\quad
\left\langle\left(\sum_{k=1}^n \frac{\beta^{\left(i\right)}_k}{k^{\left(i-1\right)/6}}%
\right)^2\right\rangle\sim n^{1-2\left(i-1\right)/6},
\end{equation*}
because $\left\langle \beta^{\left(i\right)}_k\right\rangle=0$, and $\left\langle
\beta^{\left(i\right)}_k\beta^{\left(i\right)}_{k^{\prime}}\right\rangle=0$ for $k\not=k^{\prime}$, by
 Theorem~\ref{propo_2}(ii). Hence, since for $i=2,3$,
$n^{1-2\left(i-1\right)/6}<< n\sim \left\langle \xi_n^2\right\rangle\sim\left\langle \|v_n\|^6\right\rangle$,
we conclude that, with the above condition on $\beta^{\left(i\right)}_k$, these
neglected terms
do indeed contribute a lower order correction to (\ref{eq:vnbehaviour}).
The neglected term of order $\|v_n\|^{-3}$ is also of zero average, and
therefore treated in the same way.
Unlike the first three terms, the error in (\ref{eq:v3rw}) that is
of order $\|v_n\|^{-4}=\mathrm{O}\left(n^{-2/3}\right)$,
need not be of zero average; however after summation over $n$ it
yields a contribution of order $n^{1/3}<<n^{1/2}\sim \xi_n$, and can
therefore also be neglected. Theorem~\ref{propo_2} therefore
implies that all neglected terms in (\ref{eq:v3rw}) provide lower order contributions to the
asymptotics of $\|v_n\|$. Note the crucial role of the retained term
in (\ref{eq:xirw}) involving $\gamma$, which contributes a
term of exactly the same order as the dominant diffusive
term $\epsilon_n=\beta_n^{\left(1\right)}/D$.

\begin{figure}[tbp]
\begin{center}
\includegraphics[width=11cm, keepaspectratio]{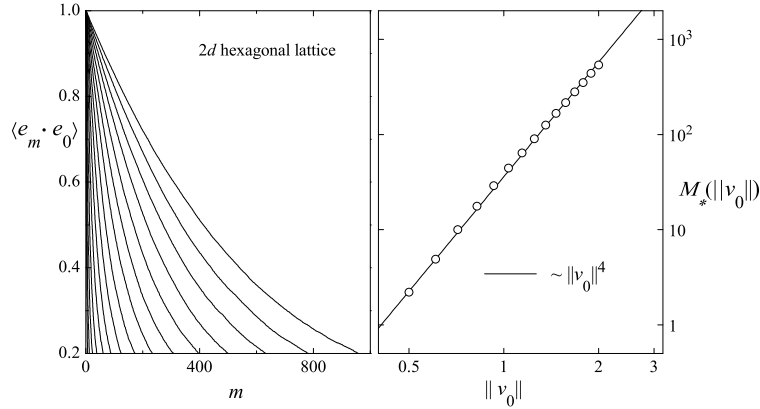}
\end{center}
\caption{On the left, the correlation function
$\left\langle e_m\cdot e_0\right\rangle $ is plotted as a function of $m$
for a set of
fifteen initial speeds $\|v_0\|$ lying in the
range 0.5 to 2.0. On the right,
a numerical estimate of $M_*\left(\|v_0\|\right)$ obtained from the initial slopes
of the data in the left panel are plotted as a function of $\|v_0\|$.}
\label{fig:turning}
\end{figure}

We now derive the asymptotic behavior of $\| y_n\|$ and $\|y\left(\tau\right)\|$
(see (\ref{eq:yn}) and (\ref{eq:ytau}) below). We first consider the
case with $d>1$, which clearly depends on how much the particle's path deviates from
a straight line, i.e., on how much and how quickly it turns. In
particular, we need to analyze the third equation in (\ref{eq:finalrw}). For
that purpose, we will first study the evolution of the unit vectors $e_n$,
which execute a random walk on the unit $\left(d-1\right)$-sphere.
Note that, for all $n$, as a result of Theorem~\ref{propo_2}(i) and the
observation that $e_n$ is independent of $\kappa_n$, the step $\delta_n$
of the walk in $e_n$, defined in  (\ref{eq:enplus1}), has a mean that
satisfies
$ \left\langle\delta_n\right\rangle=\mathrm{O}\left(\|v_n\|^{-4}\right)$. On the other hand, the
magnitude of the step is
\begin{equation*}
\|\delta_n\|=\frac{\|\alpha^{\left(1\right)}_n\|}{\|v_n\|^2}+\mathrm{O}%
\left(\|v_n\|^{-3}\right)=\|\delta_n^\perp\|. 
\end{equation*}
Given that the particle has high speed $\| v_n\|$ at the $n$th collision, we
now wish to compute how many collisions $m$ it takes for the particle's
direction to change by a macroscopic amount. For that purpose, we compute
the conditional expectation
\begin{equation*}
\left\langle\|e_{n+m}-e_n\|^2\right\rangle=\sum_{k=0}^{m-1}\sum_{k^{\prime}=0}^{m-1}%
\left\langle \delta_{n+k}\cdot\delta_{n+k^{\prime}}\right\rangle.
\end{equation*}
We will suppose $m$ satisfies $  m<<N_*\left(\xi_n\right)\sim {\|v_n\|^6}\sim n$,
so that (\ref{eq:nstarxi}) implies $\|v_{n+m}\|\sim \|v_n\|$; we will
therefore approximate $\|v_{n+k}\|$ by $\|v_n\|$. It then follows from
Theorem~\ref{propo_2}(i) that, for all $k$,
\begin{equation*}
\left\langle \delta_{n+k}\cdot\delta_{n+k}\right\rangle=\frac{\left\langle
\|\alpha_{n+k}^{\left(1\right)}\|^2\right\rangle}{\|v_n\|^{4}}+\mathrm{O}\left(\|v_n\|^{-5}\right).
\end{equation*}
For the off-diagonal terms, we note that for $k>k'$,
\begin{eqnarray*}
\left\langle \delta_{n+k}\cdot\delta_{n+k'}\right\rangle
&=& \left\langle \delta_{n+k}^\perp \cdot \delta_{n+k'}^\perp\right\rangle
+ \left\langle \mu_{n+k}e_{n+k}\cdot \delta_{n+k'}^\perp\right\rangle
+ \\
&\ &\hspace{2cm}\left\langle \delta_{n+k}^\perp\cdot \mu_{n+k'}e_{n+k'}\right\rangle+ \mathrm{O}\left(\|v_n\|^{-8}\right).
\end{eqnarray*}
In addition, the rotational invariance of the system implies that for
a given $e_{n+k}$, the vector
$\overline{\delta_{n+k}^\perp}$ vanishes (see (\ref{eq:overlinedef}) for definition of the barred average).
Hence, if $k > k'$
\begin{equation*}
\left\langle \delta_{n+k}^\perp\cdot\delta_{n+k'}^\perp\right\rangle = 0
=\left\langle \delta_{n+k}^\perp\cdot \mu_{n+k'}e_{n+k'}\right\rangle .
\end{equation*}
On the other hand, writing that $e_{n+k} = e_{n+k'+1} + \Delta_k$,
rotational invariance also implies that the conditional
expectation of $\Delta_k$ given
$e_{n+k'+1}$ is a vector $\nu_k e_{n+k'+1}$
of length $|\nu_k|\leq 2$. Hence,
$$
\left\langle\delta_{n+k'}^\perp\cdot \mu_{n+k}e_{n+k}\right\rangle
=\left\langle\delta_{n+k'}^\perp\cdot \mu_{n+k}e_{n+k'+1}\right\rangle
+ \left\langle\delta_{n+k'}^\perp\cdot \mu_{n+k}\nu_k e_{n+k'+1}\right\rangle,
$$
and
\begin{eqnarray*}
|\left\langle\delta_{n+k'}^\perp\cdot \mu_{n+k}e_{n+k}\right\rangle |
&\leq & 3\left\langle |\delta_{n+k'}^\perp\cdot e_{n+k'+1}||\mu_{n+k}|\right\rangle \\
& \leq & 3\left\langle\left|\delta_{n+k'}^\perp\cdot \left[e_{n+k'}
+\delta_{n+k'}\right]\right||\mu_{n+k}|\right\rangle\\
&\leq &\frac{3}{\|v_n\|^4}\|\delta_{n+k'}^\perp\|^2  = \mathrm{O}\left(\|v_n\|^{-8}\right).
\end{eqnarray*}
Consequently,
\begin{equation*}
\left\langle\|e_{n+m}-e_n\|^2\right\rangle=m\frac{\left\langle \|\alpha_{0}^{\left(1\right)}\|^2\right\rangle%
}{\|v_n\|^{4}} +m\mathrm{O}\left(\|v_n\|^{-5}\right) + m^2 \mathrm{O}\left(\|v_n\|^{-8}\right).
\end{equation*}
Consequently, provided
\begin{equation}  \label{eq:mstarn}
m = M_*\left(\|v_n\|\right)\sim\|v_n\|^4\sim n^{2/3}<<n
\end{equation}
we find $\left\langle\|e_{n+m}-e_n\|^2\right\rangle\sim 1$.
This shows that after $M_*(\|v_n\|)$ collisions, and aside from accidental cancelations between the diagonal and off-diagonal terms,  the particle turns through a macroscopic angle with the unit vectors
$e_{n+m}$  covering the unit sphere.
In Figure~\ref{fig:turning} we display values of $M_*(\|v_0\|)$
obtained from a numerical study of
the decay of the correlation function $\left\langle e_n\cdot e_0\right\rangle$ in
the $2d$ numerical model described in Section~\ref{s:num}. The observed power law behavior agrees with the
one predicted by the random walk analysis above.

We now analyze the asymptotic behavior of $\| y_n\|$. For particles that start off with an initial speed $\|v_0\|$, it
takes typically $M_1={\|v_0\|^4}$ collisions to acquire a
random direction of motion. We can then define
recursively
\begin{equation}  \label{eq:Nk}
M_{k+1}=M_k + M_k^{\frac{2}{3}},
\end{equation}
from which one readily finds that $ M_k\sim k^{3}.$
The $M_k$ can be interpreted by remarking that, when $n=M_k$, the particle's
velocity has ``turned'', i.e., changed direction by a macroscopic amount, on
average $k$ times, while the trajectory along the sequence
of $m << M_{k+1}-M_{k}$ collisions between $M_k$ and $M_{k+1}$ largely
follows a more or less straight path. We use this picture to
approximately compute $y_{M_k}$ by writing
\begin{equation}\label{eq:etastarwalk}
y_{M_{k+1}}=y_{M_k}+\eta_*\left(M_{k+1}-M_k\right)e_{M_k}.
\end{equation}
This is a rough estimate, but the idea is that, on average, the particles go
straight for about $M_{k+1}-M_k$ steps in the direction $e_{M_k}$ without
turning. In view of (\ref{eq:mstarn}) and (\ref{eq:Nk}) we
can now think of these successive directions $e_{M_k}$ as
randomly and independently chosen on the sphere, so that
(\ref{eq:etastarwalk}) describes a
random walk on a larger length scale, having independent steps of order
$\eta_*\left(M_{k+1}-M_k\right) \sim k^{2}. $ This yields
\begin{equation*}
\left\langle\| y_{M_k}\|^2\right\rangle\sim \sum_{\ell=1}^k \ell^{4}\sim k^{5}\sim
M_k^{5/3}.
\end{equation*}
Interpolating between the $M_k$ then allows one to write
\begin{equation}  \label{eq:yn}
\left\langle\| y_{n}\|^2\right\rangle\sim n^{5/3}.
\end{equation}
Note that, together with (\ref{eq:taun}), this finally gives the result
\begin{equation}  \label{eq:ytau}
\left\langle\| y\left(\tau\right)\|\right\rangle \sim \tau.
\end{equation}
The motion of the particles is therefore ballistic in the sense that
$\|y\left(\tau\right)\|/\tau$, which describes the rate at which
the particle's
distance from the origin grows, is finite on average. Note, however,
that the averaged instantaneous speed grows as $\tau^{1/5}$, as
shown above. The particles therefore speed up, but turn while traveling,
which decreases the rate at which they move away from the origin.
The results of our numerical calculations for $d=2$, presented
in Figure~\ref{fig:figureq2},
and described in detail in Section~\ref{s:num} are in agreement with
the results of the random walk analysis outlined above.

Finally, we briefly treat the situation for $d=1$. For this case,
the particle cannot
progressively change its direction, and so the analysis presented above
does not apply. Indeed, in $1d$ a direction change implies
a complete reversal in its direction of motion; but this can
happen only if the particle encounters a stretch of
scatterers that causes
it to completely decelerate first. Computations
similar to the previous ones show this cannot occur on a time scale
shorter than $M_*\left(\|v_0\|\right)\sim \|v_0\|^6$,
which is the same scale on which, as we have shown above,
the particle accelerates. On the other hand, for all times, we
have the obvious upper bound
\begin{equation}  \label{eq:bound}
\left\langle \|y\left(\tau \right)\|\right\rangle \leq%
 \int_0^{\tau} \rd s\ \left\langle \|v\left(\tau\right)\|\right\rangle \sim \tau^{6/5}.
\end{equation}
It is clear, then, that for time scales over which most of
the particles in the
ensemble have not reversed direction
\begin{equation}\label{eq:yn1}
\left\langle \|y_n\|\right\rangle \sim n,
\end{equation}
which along with (\ref{eq:taun})
implies $\left\langle \|y\left(\tau \right)\|\right\rangle \sim \tau^{6/5}$,
i.e., it saturates the upper bound (\ref{eq:bound}).
At longer times, the distribution of
times for the random walk (\ref{eq:xirw}) to return to the origin at
$\xi =0$, which governs events at which the velocity reverses, may
alter the asymptotics. If this happens, it does so at times longer
than we have been able to investigate numerically. Indeed, up to the times
investigated in our numerical calculations the bound (\ref{eq:bound})
appears to accurately describe the asymptotic properties of the
growth of $y\left(\tau\right)$.

\section{Non-gradient force fields}
\label{s:nongradient}

When the force field $g$ does not derive from a potential $W$, we
suppose the distribution $\nu$ of the coupling constant $c$ is
centered
\begin{equation}  \label{eq:centeredc}
\int c\ \mathrm{d} \nu\left(c\right)=0,
\end{equation}
so that the mean force vanishes at each point $%
y\in\mathbb{R}^d$.
From (\ref{eq:deltaEexpansion}), we have
\begin{equation}  \label{eq:newbessel}
\|v_{n+1}\|^2=\|v_n\|^2+2\beta_n^{\left(0\right)} + \frac{2\beta_n^{\left(1\right)}}{\|v_n\|}+%
\frac{2\beta_n^{\left(2\right)}}{\|v_n\|^2} +\mathrm{O}\left(\|v_n\|^{-3}\right).
\end{equation}
Similar to Theorem~\ref{propo_2}, we now have the following Theorem, the proof of
which also appears in the Appendix:
\begin{theorem}
\label{propo_3} Suppose Hypothesis \ref{hyp_Wtilde} and (\ref{eq:centeredc})
hold. Then:\newline
(i) For all unit vectors $e\in\mathbb{R}^d$,
$
\overline{\alpha^{\left(1\right)}\left(e\right)}=0=\overline{\alpha^{\left(2\right)}\left(e\right)}.
$
Moreover, for all $v\in\mathbb{R}^d$,
\begin{equation}  \label{eq:energyaverage2}
\overline{\Delta E\left(v\right)}=\frac{B^{\prime}}{\|v\|^2} +\mathrm{O}%
\left(\|v\|^{-3}\right),\quad \overline{\Delta E\left(v\right)^2}={{D'}^2} +\mathrm{O}%
\left(\|v\|^{-1}\right),
\end{equation}
where
$
B^{\prime}=\frac{\left(d-1\right)}{2}{D'}^{2},
$
with
\begin{eqnarray*}
{D'}^2&=&\frac{\overline{c^2}}{C_d}\int_{T^m}{\mathrm{d} \phi}
\int_{\mathbb{R}^{2d}} \mathrm{d} y_0\mathrm{d} y_0^{\prime}
\|y_0-y_0^{\prime}\|^{-\left(1+d\right)}\\
&\ &\qquad\qquad\times\left(y_0-y_0^{\prime}\right)
\cdot g\left(y_0,\phi\right) \left(y_0-y_0^{\prime}\right)\cdot g\left(y_0^{\prime},\phi\right)
\geq 0.
\end{eqnarray*}
In particular, for all unit vectors $e\in\mathbb{R}^d$ and for $\ell=0,1$,
\begin{equation}  \label{eq:betaaverage2}
\overline{\beta^{\left(\ell\right)}\left(e\right)}=0,\quad B^{\prime}=\overline{\beta^{\left(2\right)}\left(e\right)} \
\mathrm{and}\ {D'}^2=\overline{ \left(\beta^{\left(0\right)}\left(e\right)\right)^2}\geq 0.
\end{equation}
$D^{\prime}>0$ if and only if $\beta^{\left(0\right)}\left(e, \kappa\right)$ does not vanish
identically, which implies $g$ is not a gradient vector field.\newline
(ii) Let $v_n$ be the random process defined by (\ref{eq:velrw}) and $%
e_n=v_n/\|v_n\|$. Let, for $\ell\in\mathbb{N}$,
$
\beta_n^{\left(\ell\right)}=\beta^{\left(\ell\right)}\left(e_n, \kappa_n\right).
$
Then one has, for all $n\not=n^{\prime}\in \mathbb{N}$, for all $0\leq
\ell\leq\ell^{\prime}\leq 1$,
\begin{equation}
\left.
\begin{array}{l}
\label{eq:betaellindep2}
\left\langle\beta^{\left(2\right)}_n\right\rangle-B^{\prime}=0=\left\langle%
\beta_n^{\left(\ell\right)}\right\rangle  \\
\left\langle
\beta_n^{\left(\ell\right)}\beta_{n^{\prime}}^{\left(\ell^{\prime}\right)}\right\rangle=0=\left\langle
\beta_n^{\left(\ell\right)}\left(\beta_{n^{\prime}}^{\left(2\right)}-B^{\prime}\right)\right\rangle=\left\langle\left(%
\beta_{n}^{\left(2\right)}-B^{\prime}\right)\left(\beta_{n^{\prime}}^{\left(2\right)}-B^{\prime}\right)\right\rangle.
\end{array}
\right\}
\end{equation}
Moreover, $\left\langle \left(\beta^{\left(2\right)}_n\right)^2\right\rangle$ and $\left\langle\beta^{\left(\ell\right)}_n%
\beta^{\left(2\right)}_n\right\rangle$ are independent of $n$.
\end{theorem}

\begin{remark}
Whether the force does or does not depend on time plays no role in this
result, contrary to what happens in Theorem~\ref{propo_2}. In other words,
when a force field is not a gradient field, the dominant behavior of the
energy transfer to a particle is not affected by whether it depends on time
or not. In particular, the coefficients $B^{\prime}$ and $D^{\prime}$ do not
involve time derivatives of the force, as do $B$ and $D$.
\end{remark}

We now analyze the asymptotic behavior of
the velocity and the position of the particle, as in Section~\ref{s:gradient}.
From (\ref{eq:newbessel}), neglecting the subdominant terms as in
(\ref{eq:v3rw})-(\ref{eq:normalize}), we find
\begin{equation*}
\Delta \xi_n^{\prime}=\epsilon_n^{\prime}
+\frac{\gamma^{\prime}}{\xi^{\prime}_n},\quad \mathrm{where}\quad
\xi_n^{\prime}=\frac{\|v_n\|^2}{2D^{\prime}}, \quad \epsilon_n^{\prime}=%
\frac{\beta_n^{\left(0\right)}}{D^{\prime}},\quad \mathrm{and}\quad
\gamma^{\prime}=\frac14\left(d-1\right).
\end{equation*}
Note that $\gamma^{\prime}\geq 0$ in all dimensions, so that from
the analysis of (\ref{eq:xirw}) it follows that
$
\left\langle \|v_n\|^k\right\rangle\sim n^{k/4}.
$
Using this in (\ref{eq:finalrw}) yields
\begin{equation}  \label{eq:diffvtnow}
\left\langle \tau_n\right\rangle \sim \sum_{\ell=0}^n \frac1{\ell^{1/4}}\sim n^{3/4},
\quad \mathrm{and} \quad
\left\langle \|v\left(\tau\right)\|\right\rangle\sim \tau^{1/3}.
\end{equation}
This is proven rigorously in \cite{dk} for a time-independent,
non-gradient force field of the
type (\ref{eq:scatforce}) and (\ref{eq:fN}), in $d\geq 4$,  under
suitable additional technical conditions on $g$ and the distribution of the
scattering centers.

We now show what this implies for the asymptotic behavior
of $\|y\left(\tau\right)\|$.
First, the short time scale $N_*^{\prime}\left(\xi_0^{\prime}\right)$ is now
$N_*\left(\xi_0^{\prime}\right)\sim {\xi_0'}^2\sim n\sim \|v_0\|^4.$
Then, from (\ref{eq:enplus1}) we find
\begin{equation*}
e_{n+1}=e_n+\frac{\alpha_n^{\left(1\right)}-\left(\alpha_n^{\left(1\right)}\cdot e_n\right)e_n}{%
\|v_n\|^2}+\mathrm{O}\left(\|v_n\|^{-3}\right).
\end{equation*}
Consequently, $\left\langle\|e_{n+m}-e_n\|^2\right\rangle\sim m/\|v_n\|^4 \sim m/n.$
Thus, the particle now turns over a macroscopic angle after $M_*\left(\|v_n\|\right)\sim
\|v_n\|^4\sim n$ collisions, many more than for force fields
deriving from a potential (see (\ref{eq:mstarn})
and of the same order as the number $N_*\left(\|v_n\|\right)\sim n$ of
collisions it needs to accelerate significantly. This is simply due to
the fact that the particle is much faster, since $\|v_n\|\sim n^{1/4}$
rather than $\|v_n\|\sim n^{1/6}$, and harder to deflect.
This reflects itself in the
asymptotic behavior of $\|y\left(\tau\right)\|$ as follows. We define as before
$M_1=\|v_0\|^4$,  $M_{k+1}=M_k+M_k$, so that $M_k\sim 2^k$,
and
$
y_{M_{k+1}}=y_{M_k}+ \eta_*\left(M_{k+1}-M_k\right)e_{M_k},
$
which integrates to $ \left\langle \|y_{M_k}\|\right\rangle \sim M_k, $ yielding
\begin{equation}  \label{eq:diffqtnow}
\left\langle \|y\left(\tau\right)\|\right\rangle \sim \tau^{4/3},
\end{equation}
independent of the dimension $d$ of the ambient space.


\section{Homogeneous random fields}\label{s:gfields}
As we now briefly indicate, the analysis of the previous sections can be adapted to the case where the force field is not of the form (\ref{eq:rescnewton}), but is a time and space homogeneous random vector field satisfying
$$
\left\langle G\left(y,\tau\right)\right\rangle=0,\qquad \left\langle G\left(y,\tau\right)G\left(y',\tau'\right)\right\rangle =C\left(y-y', \tau-\tau'\right).
$$
Note that $C$ is a matrix-valued function, which we assume decays quickly in its
spatial variable, but not necessarily in its temporal variable.

In this situation, also, we expect the asymptotic motion of the particle to
be well described by a random walk similar to the one in (\ref{eq:finalrw}), where
now the
time step $\Delta \tau_n$ is determined by the time the particle needs to
travel through a distance $\eta_*$ equal to several times the correlation length
(which equals $1$ in the rescaled units used here) of the force field:
\begin{equation}\label{eq:finalrwhrf}
\left. \begin{array}{lll}
v_{n+1}&=&v_n+R\left(y_n,v_n,\tau_n,\Delta \tau_n\right)\\
\tau_{n+1}&=&\tau_n+\frac{\eta_*}{\| v_{n}\|},\ \eta_*\geq 1\\
y_{n+1}&=&y_n+ \eta_* e_{n}.
\end{array}\right\}
\end{equation}
Here, $R\left(y_n,v_n,\tau_n,\Delta \tau_n\right)$ is the momentum change experienced
by a particle that, after arriving at $y_n$ at
time $\tau_n$ with momentum $v_n$,
travels for a time $\Delta \tau_n$.

We consider first the case in which $G =-\nabla W$ is a random gradient field such that
\begin{equation}\label{eq:potfluct}
\left\langle W\left(y,\tau\right)\right\rangle=0,\quad \left\langle W\left(y,\tau\right)W\left(y',\tau'\right)\right\rangle=K\left(y-y',\tau-\tau'\right),
\end{equation}
where $K$ is a function of compact support in $\mathbb{B}\left(0,1\right)$ belonging to
$\mathcal{C}^5\left(\mathbb{R}^d\times\mathbb{R},\mathbb{R}\right),$ that is rotationally invariant and
even in its temporal variable.

To study the asymptotic
behavior of $v_n$ in (\ref{eq:finalrwhrf}) we first need, as in the previous sections,
to understand the asymptotic behavior of
\begin{equation}\label{eq:kinrw}
\|v_n\|^2 = \|v_0\|^2 + \sum_{k=0}^{n-1} \Delta\|v_k\|^2
=\|v_0\|^2 + \sum_{k=0}^{n-1} 2\Delta H_k - 2\left(W_n - W_0\right),
\end{equation}
where $H_k = H\left(y_k,v_k,\tau_k\right)= \|v_k\|^2/2 + W_k$, and $W_k=W\left(y_k,\tau_k\right)$.
Introducing
$$
\Delta H\left(y,v,\tau,\Delta\tau\right)
= H\left(y\left(\tau+\Delta\tau\right),v\left(\tau+\Delta\tau\right),\tau+\Delta\tau\right)-H\left(y,v,\tau\right)
$$
we find
$$\Delta H\left(y,v,\tau,\frac{\eta_*}{\left\|v\right\|}\right)
=\Delta H_I\left(y,v,\tau,\frac{\eta_*}{\left\|v\right\|}\right)
+\Delta H_{II}\left(y,v,\tau,\frac{\eta_*}{\left\|v\right\|}\right)
+\mathrm{O}\left(\left\|v\right\|^{-5}\right),$$
where
$$\Delta
H_I\left(y,v,\tau,\frac{\eta_*}{\left\|v\right\|}\right)
=\frac{\eta_*}{\left\|v\right\|}\int_0^1\rd\lambda\partial_\tau
W\left(y+\eta_*\lambda e,\tau+\frac{\eta_*\lambda}{\left\|v\right\|}\right)$$
and
\begin{eqnarray*}
\Delta H_{II}\left(y,v,\tau,\frac{\eta_*}{\left\|v\right\|}\right)
&=&-\frac{\eta_*^3}{\left\|v\right\|^3}\int_0^1\rd\lambda\nabla\partial_\tau
W\left(y+\eta_*\lambda
  e,\tau+\frac{\eta_*\lambda}{\left\|v\right\|}\right)
\\ & & \  \cdot\int_0^\lambda\rd\lambda'\int_0^{\lambda'}\rd\lambda''\nabla
W\left(y+\eta_*\lambda''
  e,\tau+\frac{\eta_*\lambda''}{\left\|v\right\|}\right).
\end{eqnarray*}
We then have the same kind of result as in Theorem \ref{propo_2}, the proof
of which is immediate:
\begin{proposition}
Under the above conditions,
$\left\langle\alpha^{\left(1\right)}\right\rangle=0=\left\langle\alpha^{\left(2\right)}\right\rangle$
and for all $v\in\mathbb{R}^d,$
$$\left\langle\Delta
H\left(v\right)\right\rangle=\frac{\tilde B}{\left\|v\right\|^{4}}+\mathrm{O}\left(\left\|v\right\|^{-5}\right),\
\left\langle\left(\Delta
  H\left(v\right)\right)^2\right\rangle=\frac{\tilde D^2}{\left\|v\right\|^{2}}+\mathrm{O}\left(\left\|v\right\|^{-3}\right),$$
where
$$\tilde B=\left(d-3\right)\eta_*K^{\left(0\right)}-2\left(d-4\right)K^{\left(1\right)},\
\tilde D^2=2\left(\eta_*K^{\left(0\right)}-K^{\left(1\right)}\right),$$
and
$$K^{\left(0\right)}=\int_0^1\rd\mu\left(-\partial_t^2K\left(\mu
  e,0\right)\right),\ K^{\left(1\right)}=\int_0^1\rd\mu\left(-\mu\partial_t^2K\left(\mu
  e,0\right)\right).$$
\end{proposition}
\begin{proof}
Noting that $\left\langle\Delta H_I\left(v\right)\right\rangle=0,$ we find
$$
\begin{array}{l}
\left\langle\Delta
H\left(v\right)\right\rangle=\left\langle\Delta
H_{II}\left(v\right)\right\rangle
\\ \quad=\frac{\eta_*^3}{\left\|v\right\|^3}\int_0^1\rd\lambda\int_0^{\lambda}\rd\lambda''\left(\lambda-\lambda''\right)\left(\Delta\partial_t
K\right)\left(\eta_*\left(\lambda-\lambda''\right)e,\frac{\eta_*\left(\lambda-\lambda''\right)}{\left\|v\right\|}\right)
\\ \quad=\frac{\eta_*^4}{\left\|v\right\|^4}\int_0^1\rd\lambda\int_0^{\lambda}\rd\lambda''\left(\lambda-\lambda''\right)^2\left(\Delta\partial_t^2
K\right)\left(\eta_*\left(\lambda-\lambda''\right)e,0\right)+\mathrm{O}\left(\left\|v\right\|^{-5}\right)
\\
\quad=\frac{\eta_*^4}{\left\|v\right\|^4}\int_0^1\rd\lambda\left(1-\lambda\right)\lambda^2\left(\Delta\partial_t^2
K\right)\left(\eta_*\lambda e,0\right)+\mathrm{O}\left(\left\|v\right\|^{-5}\right).
\end{array}
$$
Using the rotational invariance of $\Delta\partial_t^2K\left(\cdot,0\right)$, and integrating
by parts, we obtain the above expression for $\tilde B.$
\end{proof}

Scaling $\|v_n\|^2$ by $\left(s/n\right)^{1/3}$ in (\ref{eq:kinrw}) and taking $n$ to infinity, one finds
that the limiting process $Z_\sigma$ satisfies the stochastic differential equation
$$
\rd Z_\sigma = \frac23 \frac{dB_\sigma}{\sqrt{Z_\sigma}}+\frac23\left(\gamma-\frac16\right)\frac{\rd \sigma}{Z_\sigma^2},\quad \gamma=\frac{1}{3}\left(\frac{\tilde B}{\tilde D^2}+\frac{1}{2}\right).
$$
It then follows from the It\^o formula that $Y_\sigma=Z_\sigma^3$ satisfies the stochastic differential equation of the square of the Bessel process \cite{ry} of dimension $\delta = 2\gamma + 1$. By taking $\eta_*$ sufficiently large we can  make $\gamma$ arbitrarily close to $\left(d-2\right)/6$.
The analysis of the random walk is therefore entirely analogous to the one in
Section~\ref{s:gradient}, yielding in particular the same power laws for the growth
of $\left\langle v^2\left(\tau\right)\right\rangle$ and $\left\langle y^2\left(\tau\right)\right\rangle$ as in
(\ref{eq:diffvtw}) and (\ref{eq:ytau}).

In the case that $G$ is not a gradient field
we still assume it to be rotationally invariant and reflection symmetric.
This implies that
there exist functions $\Lambda_1$ and $\Lambda_2$ such that the
correlation function is of the form
$$C\left(y,\tau\right)=\Lambda_1\left(\left\|y\right\|,\tau\right)\mathbb{P}_y+\Lambda_2\left(\left\|y\right\|,\tau\right)\mathbb{P}_y^{\perp},$$
where $\mathbb{P}_y$ is the orthogonal projector along the direction of the vector $y$ and
$\mathbb{P}_y^{\perp}+\mathbb{P}_y=I_d$.
We in addition assume that $\Lambda_1$ and $\Lambda_2$
are $\mathcal{C}^2$ functions that decay fast in their spatial variable,
and that for all $\tau\in\mathbb{R}, \Lambda_1\left(\cdot,\tau\right)$
and $\Lambda_2\left(\cdot,\tau\right)$ are compactly supported  in $[0,1].$
Under these assumption, we then prove an analogue of Theorem~\ref{propo_3}:
\begin{proposition}
Under the conditions stated above
$\left\langle\alpha^{\left(1\right)}\right\rangle=0=\left\langle\alpha^{\left(2\right)}\right\rangle$, and
for all $v\in\mathbb{R}^d,$
$$\left\langle\Delta E\left(v\right)\right\rangle=\frac{\tilde B'}{\left\|v\right\|^2}+\mathrm{O}\left(\left\|v\right\|^{-3}\right),\qquad \left\langle\left(\Delta E\left(v\right)\right)^2\right\rangle=\tilde D'^2+\mathrm{O}\left(\left\|v\right\|^{-1}\right),$$
where
$$\tilde B'=\eta_*\left(d-1\right)K'^{\left(0\right)}-\left(d-2\right)K'^{\left(1\right)},\qquad
\tilde D'^2=2\left(\eta_*K'^{\left(0\right)}-K'^{\left(1\right)}\right)>0$$
and
$$K'^{\left(0\right)}=\int_0^1\rd\mu\Lambda_1\left(\mu,0\right),\qquad K'^{\left(1\right)}=\int_0^1\rd\mu\mu\Lambda_1\left(\mu,0\right).$$
\end{proposition}
\begin{proof}
A computation of
$R\left(y,v,\tau,\eta_*/\left\|v\right\|\right)=\int_{\tau}^{\tau+\eta_*/\left\|v\right\|}G\left(y\left(\tau'\right),\tau'\right) \rd\tau'$  to second order in perturbation theory gives
$$R\left(y,v,\tau,\eta_*/\left\|v\right\|\right)=R_I\left(y,v,\tau,\eta_*/\left\|v\right\|\right)+R_{II}\left(y,v,\tau,\eta_*/\left\|v\right\|\right)$$
with
$$R_I\left(y,v,\tau,\eta_*/\left\|v\right\|\right)=\frac{\eta_*}{\left\|v\right\|}\int_0^1\rd\lambda
G\left(y+\eta_*\lambda e,\tau+\frac{\eta_*\lambda}{\left\|v\right\|}\right)$$
and
\begin{eqnarray*}
R_{II}\left(y,v,\tau,\eta_*/\left\|v\right\|\right)&=&\frac{\eta_*^3}{\left\|v\right\|^3}\int_0^1\rd\lambda\int_0^\lambda\rd\lambda''\left(\lambda-\lambda''\right)\times\\
&\ &\left(G\left(y+\eta_*\lambda''
  e,\tau\right)\cdot\nabla\right)G\left(y+\eta_*\lambda
e,\tau\right)+\mathrm{O}\left(\left\|v\right\|^{-4}\right).
\end{eqnarray*}
Hence, $\left\langle\alpha^{\left(1\right)} \right\rangle$ and $\left\langle\alpha^{\left(2\right)}\right\rangle,$
and consequently $\left\langle\beta^{\left(0\right)}\right\rangle$ and $\left\langle\beta^{\left(1\right)}\right\rangle,$
vanish.
Using $w\cdot\mathbb{P}_y\left(v\right)=\left(v\cdot y\right)\left(w\cdot y\right)/y^2$ then yields

\begin{eqnarray*}
\frac{1}{2}\left\langle\alpha^{\left(1\right)}\cdot\alpha^{\left(1\right)}\right\rangle&=&\\
&\ &\hspace{-2cm}=\frac{\eta_*^2}{2}\int_0^1\rd\lambda\int_0^1\rd\lambda''
\left(\Lambda_1\left(L\left|\lambda-\lambda''\right|,0\right)+\left(d-1\right)\Lambda_2\left(\eta_*\left|\lambda-\lambda''\right|,0\right)\right)
\\
&\ &\hspace{-2cm}=\eta_*^2\int_0^1\rd\lambda\left(1-\lambda\right)
\left(\Lambda_1\left(\eta_*\lambda,0\right)+\left(d-1\right)\Lambda_2\left(\eta_*\lambda,0\right)\right),\quad \mathrm{and}
\end{eqnarray*}
\begin{eqnarray*}
\left\langle\alpha^{\left(3\right)}\cdot e\right\rangle&=&\\
&\ &\hspace{-2.5cm}=\eta_*^2\int_0^1\rd\lambda\left(1-\lambda\right)\left(\eta_*\lambda\Lambda_1'\left(\lambda,0\right)
+\left(d-1\right)\left(\Lambda_1\left(\lambda,0\right)-\Lambda_2\left(\lambda,0\right)\right)\right)
\\  
&\ &\hspace{-2.5cm}=\eta_*^2\int_0^1\rd\lambda\left(\left(\left(d-2\right)-
\lambda\left(d-3\right)\right)\Lambda_1\left(\eta_*\lambda,0\right)-\left(d-1\right)\left(1-\lambda\right)\Lambda_2\left(\eta_*\lambda,0\right)\right).
\end{eqnarray*}
Adding the last two equations and making the change of variables $\mu=\eta_*\lambda$ yields the above expression for $\tilde B'$.
\end{proof}
Analysis of the random walk now proceeds along the lines of Section~\ref{s:nongradient}, yielding the same power laws as obtained therein.


\section{Numerical results}

\label{s:num} To verify our theoretical analysis of the motion of
a particle in random force fields presented in the previous sections, we
performed numerical calculations for a periodic array of soft
scatterers in one and two dimensions. For the two dimensional
case we employed a
hexagonal lattice, with, for $N=\left(N_1, N_2\right)\in\mathbb{Z}^2$,
$
x_N=N_1u +N_2v, \quad\mathrm{where}\quad u=\left(1,0\right),\quad v=\frac12\left(1,\sqrt3\right).
$
We focused on the case in which the force fields associated with
the scatterers were derived from a potential, taking $W$ to be
of the form
of a time-dependent, flat circular potential,
\begin{eqnarray}
W\left(y,\phi\right)&=&f\left(\phi\right)\chi\left(\frac{\|y\|}{y_*}\right),\quad y\in\mathbb{R}^d, d=1,2,
\notag
\end{eqnarray}
where $\chi\left(x\right)=1$ if $0\leq x\leq 1$ and $\chi\left(x\right)=0$ otherwise. Here the
parameter $y_*$ satisfies $\frac{\sqrt3}{4}<y_*<1/2$ to ensure the system has a finite horizon. Three different
choices were explored for the function $f$, namely,
$$
f_1\left(\phi\right)=\cos\left(2\pi\phi\right),\quad f_2\left(\phi\right)=1+\cos^2\left(2\pi\phi\right), \quad
\phi\in[0,1[,
$$
each of which leads to a time-periodic potential, and
\begin{equation*}
f_3\left(\phi\right)=f_3\left(\phi_1,\phi_2\right)=\cos\left(2\pi\phi_1\right)+\cos\left(2\pi\phi_2\right).
\end{equation*}
In the latter case, the frequency vector $\omega$ was chosen to be $%
\omega=\left(1,\sqrt2\right)$ so that the resulting potential is quasi-periodic in
time. The phases $\phi_N$ were chosen uniformly on the torus,
independently for each scatterer. Coupling constants $c_N$ were either
drawn independently from a uniform distribution on $[0, 1/2]$,
or set to a fixed value $c_N=1$, or $c_N=-1$, for all $N$.

Depending on the phase and the choice of coupling constants, each such potential
describes a centrally symmetric potential barrier or well,
whose maximum/minimum oscillates in time. For the
choice $f=f_1$ or $f_3$, any given scatterer
will sometimes act as a potential well, and at other times as a
barrier, depending
on the sign of $c_Nf_1\left(\phi_N+\tau\right)$ or of $c_Nf_3\left(\phi_N+\omega\tau\right)$ at the time $\tau$ of arrival of the
particle; on average the force at a given point in space always vanishes.
When $f=f_2$ on the other hand, and $c_N=1$ for all $N$, $c_Nf_2\left(\phi+\tau\right)$ is
always positive, yielding a lattice of oscillating potential
barriers, for which the average force at a given point in space does not
vanish. Similarly, when $f=f_2$ and $c_N=-1$ for all $N$ one obtains a
lattice of oscillating potential wells. In all cases that we studied numerically,
the system had finite horizon.

Motion of a particle through an array of such scatterers can be computed iteratively,
by using energy and angular momentum conservation at the entry and exit
of the particle from the support of the potential, and without a numerical
integration of a second order differential equation. This allows one to
compute the motion of the particle numerically for very long times, as
required to properly study the asymptotic regime.

In our calculations, each particle was initially placed
at a point randomly chosen on the boundary of the scatterer
at the origin, with an initial velocity drawn with equal
probability from all possible outward directions.
For each ensemble of initial conditions, the initial speed $\|v_0\|$ of the
particle was kept fixed (with values indicated in
the figure captions, or in the figures themselves).  Displayed
results represent averages over, typically, $10^4$ trajectories
for each initial speed.  For convenience of presentation, data
appearing as a function of time $\tau$ or collision number $n$ in the
figures presented throughout the paper represent a subset
of the data generated, evaluated at values of the independent variable
that are equally spaced on a logarithmic axis.

A general finding of both our numerical calculations and of our theoretical
analysis is that the
power law behaviors associated with stochastic
acceleration is independent of the precise form of
the potential employed; in particular it does not depend on whether
the average force vanishes or not. Thus, in the figures that appear in the
paper we have
chosen to present numerical results only for the case
where $f=f_1$ and $c_N$ is uniformly distributed in $[0,1/2]$.

For this specific model, Figure~\ref{fig:figurev2} shows the evolution of
the particle's mean kinetic
energy, both as a function of time $\tau$ and as a function of
collision number $n$. As noted in the text, one observes excellent
agreement with the power law behavior predicted by our analysis
(see (\ref{eq:asp2t}), (\ref{eq:vnbehaviour}), (\ref{eq:diffvtw})),
independent of dimension. One
also notices in this figure that the asymptotic regime is reached after an
initial period which ends after a number $N_*\left(\|v_0\|\right)$ of collisions that
grows with $\|v_0\|$. The value of $N_*\left(\|v_0\|\right)$ was computed numerically
for fifteen values of $\|v_0\|$ between $0.5$ and $2$, and the result is
presented in Figure~\ref{fig:regimechange}. The observed power law
$N_*\left(\|v_0\|\right)\sim\|v_0\|^6,$ is as predicted in
Section~\ref{s:gradient} (see equation (\ref{eq:nstarxi})).

Similarly, Figure~\ref{fig:figureq2} shows for the specific
numerical model described above,
the evolution of the particle's
mean squared displacement as a function of $\tau$ and $n$. We find that
the power laws obtained in one dimension (see (\ref{eq:asq2t1}) and
(\ref{eq:yn1})) and in two dimensions (see (\ref{eq:asq2td}) and (\ref{eq:yn})-(\ref{eq:ytau})) are
indeed different, and precisely as predicted by the analysis
of Section~\ref{s:gradient}.

 In order to obtain analytical results for a sufficiently general class of
potentials, the theoretical analysis of Section~\ref{s:gradient} assumed
scattering potentials that are smooth, which the potentials used in
the numerical calculations are clearly not. Indeed, running the
numerics for sufficiently long
times with a smooth potential would involve repeatedly solving a
second order differential
equation; this would lack precision and be too time consuming. Explicit computations
specific to the square potential, however, show that formulas
(\ref{eq:deltaEexpansion}) and (\ref{eq:Rexpansion}) remain valid,
and that their dominant terms have the same behavior as in
the analysis presented, so that our arguments go through
unaltered. This lends additional support to our claim that
it is the high energy behavior
of the energy and momentum transfer in a single scattering event that
determines the asymptotic behavior of the particle, and suggests that
the results are even more universal than is implied by our analysis.

As a closing comment we note also that in our numerical models the potentials
are rotationally invariant, and the lattices are ordered. Thus, when
$c_N$ is constant, the only randomness left in the problem is in the
initial phases $\phi_N$ of the scatterers and the initial
directions  $e_0$ of the particles. Thus, the essential randomness
necessary for the validity of our analysis arises from the dispersive
nature of the scattering event itself, which leads to
a random sequence of scattering events when evaluated along the
trajectory that the particle follows.


\appendix
\section{Proof of main theorems}\label{s:bols}

In this appendix we provide proofs of Theorems~\ref{propo_2} and
\ref{propo_3}.
We begin with some preparatory material. It
is convenient to write $\hat g\left(y,\tau\right)=Mg\left(M^{-1}y, \omega\tau+\phi\right)$, suppressing
the variables $\phi$ and $M$ from the notation.

The estimates below are all uniform in $\phi$ and $M$. Note that
when $g=-\nabla W$, then $\hat g=-\nabla \hat W$, with
$
\hat W\left(y,\tau\right)=W\left(M^{-1}y,\omega\tau+\phi\right).
$
We need to study the solutions of
\begin{equation}  \label{eq:newtonphi}
\ddot y \left(\tau^{\prime}\right)=c\hat g\left(y\left(\tau^{\prime}\right), \tau^{\prime}\right)=-c\nabla
\hat W\left(y\left(\tau^{\prime}\right),\tau^{\prime}\right), \ y\left(\tau_0\right)=y_0, \ \dot
y\left(\tau_0\right)=v_0.
\end{equation}
For any initial condition $y\left(\tau_0\right)=y_0, v\left(\tau_0\right)=v_0$, we define
\begin{equation}  \label{eq:pasymp}
v_\pm = \lim_{\tau\to\pm\infty} \dot y\left(\tau\right).
\end{equation}
That these limits exist if $\| v_0\|$ is large enough is a consequence of
the following lemma.

\begin{lemma}
\label{lem:estim_ts} Suppose Hypothesis \ref{hyp_Wtilde} holds.
Let $\tau_0\in\mathbb{R}$ and suppose $\left(y_0,v_0\right)\in\mathbb{R}^{2d}$
satisfies $\| y_0\| \leq 1/2$, $\| v_0\|^2\geq 12cg_{\mathrm{max}}$. Then there exist
unique $\tau_{\mathrm{in}}\leq \tau_0\leq \tau_{\mathrm{out}}$ so that $\|
y\left(\tau_{\mathrm{in}}\right)\| =\frac52=\| y\left(\tau_{\mathrm{out}}\right)\|$. Moreover
\begin{equation}  \label{eq:estim_ts2}
\frac{1}{\| v_0\|}\leq \min\{\tau_0-\tau_{\mathrm{in}}, \tau_{\mathrm{out}%
}-\tau_0\}\leq \max\{\tau_0-\tau_{\mathrm{in}}, \tau_{\mathrm{out}%
}-\tau_0\}\leq \frac{5}{\| v_0\|}.
\end{equation}
\end{lemma}

The lemma roughly says that any particle that is at some instant $\tau_0$
inside the region where the potential does not vanish and that has enough
kinetic energy at that moment, has entered it in the past and will leave
again in the future, spending a time of order $\frac{1}{\| v_0\|}$ to cross
it: both the upper and lower bounds in (\ref{eq:estim_ts2}) will be used in
the proof of Proposition \ref{prop:deltat} below. Note that the Lemma does
indeed imply the existence of the limits in (\ref{eq:pasymp}).

\begin{proof}
From (\ref{eq:newtonphi}),
\begin{equation}\label{eq:inteqmotion}
y\left(\tau\right)=y_0+v_0\left(\tau-\tau_0\right)+c\int_{\tau_0}^{\tau}d\tau' \int_{\tau_0}^{\tau'}d\tau''
\hat g\left(y\left(\tau''\right),\tau''\right),
\end{equation}
so that
$
Q\left(\tau-\tau_0\right)\leq\left\|y\left(\tau\right)\right\|\leq P\left(\tau-\tau_0\right),
$
where
$$
Q\left(\tau-\tau_0\right)=-cg_{\mathrm{max}}\frac{\left(\tau-\tau_0\right)^2}{2}+\| v_0\| |\tau-\tau_0|-\frac12
$$
and
$$
P\left(\tau-\tau_0\right)= cg_{\mathrm{max}}\frac{\left(\tau-\tau_0\right)^2}{2}+\| v_0\| |\tau-\tau_0|+\frac12.
$$
One checks that $Q\left(\sigma_+\right)=\frac{5}{2}=P\left(\sigma_-\right)$, with
$$
\sigma_-=\frac{\left\| v_0\right\|}{cg_{\mathrm{max}}}\left(\sqrt{1+\frac{4cg_{\mathrm{max}}}{\left\| v_0\right\|^2}}-1\right),\ \mathrm{and}\
\sigma_+=\frac{\left\| v_0\right\|}{cg_{\mathrm{max}}}\left(1-\sqrt{1-\frac{6cg_{\mathrm{max}}}{\left\| v_0\right\|^2}}\right).
$$
Note that $\sigma_-\leq \sigma_+$. Since
$
\|y\left(\tau_0\pm\sigma_-\right)\|\leq \frac52\leq \|y\left(\tau_0\pm \sigma_+\right)\|,
$
it is clear there exist $\tau_{\mathrm{in}}, \tau_{\mathrm{out}}$ satisfying
$$
\tau_0-\sigma_+\leq \tau_{\mathrm{in}}\leq \tau_0-\sigma_-\quad\mathrm{and}\quad \tau_0+\sigma_-\leq \tau_{\mathrm{out}}\leq \tau_0+\sigma_+.
$$
Uniqueness follows from the observation that $\hat g$ vanishes outside the ball of radius $1/2$ so that the trajectory can enter and leave the ball of radius $5/2$ only once.
Equation (\ref{eq:estim_ts2}) now follows from the observation that, if $0\leq x\leq A^{-1}\leq 1$, then
$$
\sqrt{1+x}-1\geq \frac12\frac{\sqrt{A}}{\sqrt{1+A}}x,\ 1-\sqrt{1-x}\leq\frac12\frac{\sqrt{A}}{\sqrt{A-1}}x.
$$
It is enough to choose $A=1$ in the first inequality and $A=2$ in the second.
\end{proof}
With $v_\pm$ from (\ref{eq:pasymp}), we define, for all $\left(y_0,
v_0,\tau_0\right)$ so that $\|v_0\|^2\geq12cg_{\mathrm{max}}$,
\begin{equation}  \label{eq:defdeltav}
\Delta v\left(v_0, y_0,\tau_0\right)=v_+-v_-,\quad \Delta K\left(v_0, y_0, \tau_0\right)=\frac{1}%
2\left(v_+^2-v_-^2\right).
\end{equation}
Note that both $\Delta v$ and $\Delta K$ are constant along trajectories:
\begin{equation}  \label{eq:trajfunction}
\Delta v\left(v_0, y_0, \tau_0\right)=\Delta v(v(\tau^{\prime}), y(\tau^{\prime}),
\tau^{\prime}),\quad \Delta K(v_0, y_0, \tau_0)=\Delta K(v(\tau^{\prime}),
y(\tau^{\prime}), \tau^{\prime}).
\end{equation}
We therefore think of them as functions on the space of all trajectories
with sufficient kinetic energy. We are interested in understanding the high
velocity behavior of $\Delta K$ and of its average over all those
trajectories that enter the support of the potential. We will see that, when
$\hat g=-\nabla \hat W$, $\Delta K\sim \| v_0\|^{-1}$
(Proposition \ref{prop:deltat}) but that the average of $\Delta K$ vanishes up to terms of order $
\|v_0\|^{-4}$ (Proposition \ref{prop:avdeltat}). In other words, the
dominant terms in $\Delta K$ vanish on average. This observation
is the essence of Theorem~\ref{propo_2}, as we will explain below.

\begin{proposition}
\label{prop:deltap} Suppose Hypothesis \ref{hyp_Wtilde} is satisfied.
Let $y_0, v_0\in\mathbb{R}^d$ and $\tau_0\in\mathbb{R}$, with $\|v_0\|^2\geq
12cg_{\mathrm{max}}$. Then
\begin{eqnarray}  \label{eq:deltav}
\Delta v\left(v_0, y_0, \tau_0\right)&=&\frac{c}{\| v_0\|}\int_{-\infty}^{+\infty} \hat
g\left(y_0+\lambda e_{0}, \tau_0\right)\ \mathrm{d}\lambda  \notag \\
&\ &\hspace{-2cm}+\frac{c}{\| v_0\|^2}\int_{-\infty}^{+\infty} \partial_\tau
\hat g\left(y_0+\lambda e_{0}, \tau_0\right)\lambda \mathrm{d}\lambda+\mathrm{O}\left(\|
v_0\|^{-3}\right).
\end{eqnarray}
The error term is uniform in $y_0, \tau_0$, $c\in[-1,1]$ and in $e_0=v_0/\|
v_0\|$.
\end{proposition}

\begin{proof}
Suppose first $\| y_0\| \leq \frac12$. It follows from Lemma \ref{lem:estim_ts} that, under
the stated condition on $\| v_0\|$, there exist unique entrance
and exit times  $\tau_{\mathrm{in}}$ and $\tau_{\mathrm{out}}$ to the ball of
radius $5/2$, with $\tau_{\mathrm{out}}-\tau_0$ and $\tau_0-\tau_{\mathrm{in}}$ of order $\|v_0\|^{-1}$.
As a result, for $\tau<\tau_{\mathrm{in}}$ and $\tau>\tau_{\mathrm{out}}$ the
particle executes a free motion with speeds $v_-$ and $v_+$, in the region
where the force $\hat g$ vanishes identically.
From (\ref{eq:newtonphi}), one now readily concludes
\begin{eqnarray*}
\Delta v\left(v_0, y_0, \tau_0\right)&=&c\int_{\tau_{\mathrm{in}}}^{\tau_\mathrm{out}}
\hat g\left(y\left(\tau\right), \tau\right) \rd \tau\\
&=&c\int_{\tau_{\mathrm{in}}}^{\tau_\mathrm{out}}
\hat g\left(y_0 +v_0\left(\tau-\tau_0\right), \tau\right) \rd \tau+\mathrm{O}\left(\|v_0\|^{-3}\right),
\end{eqnarray*}
where we used
$
\|y\left(\tau\right)-\left(y_0+v_0\left(\tau-\tau_0\right)\right)\|\leq \frac{c}{2}g_{\mathrm{max}}\left(\tau-\tau_0\right)^2,
$
which follows easily from (\ref{eq:inteqmotion}).
Let us now remark that Lemma \ref{lem:estim_ts} implies that
$$
\|y_0+v_0\left(\tau_{\mathrm{out/in}}-\tau_0\right)\|\geq \|v_0\||\tau_{\mathrm{out/in}}-\tau_0|-\frac12\geq 1/2.
$$
As a result, we can extend the $\tau$ integration to the full real axis; indeed, the  integrand vanishes for
$\tau\leq\tau_{\mathrm{in}}$ and for $\tau_{\mathrm{out}}\leq \tau$.  The change of variables $\lambda =\| v_0\| \left(\tau-\tau_0\right)$ yields
\begin{equation}\label{eq:deltav2}
 \Delta v\left(v_0, y_0, \tau_0\right)=\frac{c}{\|v_0\|}\int_{-\infty}^{+\infty}
\hat g\left(y_0 +\lambda e_0, \tau_0+\frac{\lambda}{\|v_0\|}\right) \rd \lambda +\mathrm{O}\left(\|v_0\|^{-3}\right),
\end{equation}
so that a first order Taylor expansion yields the result.

We now consider the case where $\| y_0\| >1/2$. We may assume the particle trajectory intersects the ball of radius $1/2$ centered at the origin: otherwise $\Delta v\left(v_0, y_0,\tau_0\right)=0$, and then the result stated certainly holds. Suppose therefore the trajectory intersects that ball and that $y_0\cdot v_0\leq0$. Then  there exists a unique time $\tau_*>\tau_0$ when the trajectory enters the above ball: so $y\left(\tau\right)=y_0+v_0\left(\tau-\tau_0\right)$ for all $\tau\leq\tau_*$, $\|y\left(\tau_*\right)\|=1/2$ and $y\left(\tau_*\right)\cdot v_0\leq 0$.
Clearly
$$
\Delta v\left(v_0, y_0,  \tau_0\right)=\Delta v \left(v_0, y\left(\tau_*\right), \tau_*\right),
$$
and we can apply the result of the first part of the proof to write
\begin{equation}\label{eq:5}
\Delta v\left(v_0, y_0,\tau_0\right)= \frac{c}{\| v_0\|}\int_{-\infty}^{+\infty} \hat g\left(y\left(\tau_*\right)+\lambda e_{0}, \tau_*+\frac{\lambda}{\|v_0\|}\right)\ \rd\lambda
+\mathrm{O}\left(\| v_0\|^{-3}\right).
\end{equation}
The change of variables
$$
\tilde \lambda = \lambda +\|v_0\|\left(\tau_*-\tau_0\right)
$$
transforms (\ref{eq:5}) into (\ref{eq:deltav2}), which concludes the proof. The case where $y_0\cdot v_0\geq 0$ is treated analogously.
\end{proof}

When $\hat g=-\nabla\hat W$, we need the high $\|v_0\|$ expansion of $\Delta
K$ up to order $\|v_0\|^{-4}$ obtained in the following proposition.

\begin{proposition}
\label{prop:deltat} Suppose Hypothesis~\ref{hyp_Wtilde} is satisfied and
suppose $\hat g=-\nabla\hat W$. Then, for all $v_0\in\mathbb{R}^d$ such that
$\| v_0\|^2 \geq 12cg_{\mathrm{max}}$ and for all $y_0\in\mathbb{R}^d$,
\begin{equation}  \label{eq:deltat}
\Delta K\left(v_0, y_0, \tau_0\right)=\Delta K_{I}\left(v_0, y_0, \tau_0\right)+ \Delta
K_{II}\left(v_0, y_0, \tau_0\right) +\mathrm{O}\left(\| v_0\|^{-5}\right)
\end{equation}
where
\begin{equation}  \label{eq:deltaenergyI}
\Delta K_I\left(v_0, y_0, \tau_0\right)=\frac{c}{\| v_0\|}\int_\R \mathrm{d} \lambda\
\partial_\tau \hat W\left(y_0+\lambda e_0, \tau_0+\frac{\lambda}{\| v_0\|}\right),
\end{equation}
and
\begin{eqnarray}  \label{eq:deltaenergyII}
\Delta K_{II}\left(v_0, y_0, \tau_0\right)&=&-\frac{c^2}{\| v_0\|^3}\int_{\mathbb{R}}%
\mathrm{d} \lambda \nabla \partial_\tau\hat W\left(y_0+\lambda e_0, \tau_0+\frac{%
\lambda}{\| v_0\|}\right)  \notag \\
&\ &\hspace{-1cm}\cdot \int_0^\lambda \mathrm{d} \lambda^{\prime}\int_0^{%
\lambda^{\prime}} \mathrm{d} \lambda^{\prime\prime}\nabla \hat
W\left(y_0+\lambda^{\prime\prime}e_0, \tau_0+\frac{\lambda^{\prime\prime}}{\|
v_0\|}\right).
\end{eqnarray}
The error term is uniform in $y_0, \tau_0$, and in $e_0=v_0/\| v_0\|$.
\end{proposition}

The index ``I'' or ``II'' refers to first and second order in $\hat W$, but
note that each of the corresponding contributions has an expansion in $%
\|v_0\|^{-1}$.

\begin{proof}
We first deal with the case where $\| y_0\| \leq 1/2$. As in the proof of Proposition \ref{prop:deltap}, one can integrate the equation of motion to obtain
\begin{equation*}
\Delta K\left(v_0, y_0, \tau_0\right)=-c\int_{\tau_{\mathrm{in}}}^{\tau_\mathrm{out}} \dot y\left(\tau\right)\cdot \nabla
\hat W\left(y\left(\tau\right),\tau\right) \rd \tau=
c\int_{\tau_\mathrm{in}}^{\tau_\mathrm{out}} \partial_\tau
\hat W\left(y\left(\tau\right),\tau\right) \rd \tau.
\end{equation*}
From (\ref{eq:inteqmotion}) one easily finds, for $\tau\in [\tau_{\mathrm{in}}, \tau_{\mathrm{out}}]$, that
\begin{equation}\label{eq:perttheory}
 \left.\begin{array}{lll}
\|\dot y\left(\tau\right)-v_0\|&\leq& cg_{\mathrm{max}}|\tau-\tau_0|\\
\|y\left(\tau\right)-\left(y_0+v_0\left(\tau-\tau_0\right)\right)\|&\leq& cg_{\mathrm{max}}\left(\tau-\tau_0\right)^2\\
y\left(\tau\right)&=&y_I\left(\tau\right) +\mathrm{O}\left(\|v_0\|^{-4}\right)
\end{array}\right\}
\end{equation}
where we used (\ref{eq:estim_ts2}) in the last line and where
$$
y_I\left(\tau\right)=y_0+v_0\left(\tau-\tau_0\right)-c\int_{\tau_0}^\tau \rd \tau' \int_{\tau_0}^{\tau'} \rd \tau''\ \nabla \hat W\left(y_0+v_0\left(\tau''-\tau_0\right),\tau''\right).
$$
Hence
$$
\Delta K\left(v_0, y_0, \tau_0\right)=c\int_{\tau_\mathrm{in}}^{\tau_\mathrm{out}} \partial_\tau
\hat W\left(y_I\left(\tau\right),\tau\right) \rd \tau +\mathrm{O}\left(\|v_0\|^{-5}\right).
$$
Expanding $\partial_\tau \hat W\left(y_I\left(\tau\right),\tau\right)$ around $y_0 + v_0\left(\tau - \tau_0\right)$, the result follows.
The case $\| y_0\| >1/2$ is handled as in the proof of Proposition~\ref{prop:deltap}.
\end{proof}

For the purpose of proving Theorem~\ref{propo_2}, we now turn to the
computation of the average energy change of all trajectories with a given,
sufficiently high, incoming momentum or energy, and that enter the ball of
radius $1/2$ centered at the origin. Recalling that $\hat g\left(y,\tau\right)=Mg\left(M^{-1}y,\omega\tau+\phi\right)$,
we have, for $v_0\in\mathbb{R}^d$, $b\cdot v_0=0$ and $\kappa=\left(b,
M,\phi+\omega\tau_0,c\right)$,
\begin{equation}  \label{eq:deltakdeltaE}
\Delta E\left(v_0,\kappa\right)= \Delta K\left(v_0, b-\frac12e_0,\tau_0\right).
\end{equation}
We first compute the average of ${\Delta E}\left(v_0, b, M,\phi, c\right)$ over $\phi$:
\begin{equation}  \label{eq:phiavdeltaE}
\int_{\mathbb{T}^m}\mathrm{d}\phi\ {\Delta E}\left(v_0, b, M,\phi, c\right).
\end{equation}

\begin{proposition}
\label{prop:avdeltat} Suppose Hypothesis \ref{hyp_Wtilde} is satisfied.
Then, for all $v_0\in\mathbb{R}^d$ and for all $b\in\mathbb{R}^d$, $b\cdot
v_0=0, M\in\mathrm{SO}\left(d,\mathbb{R}\right), c\in[-1,1]$,
\begin{eqnarray}  \label{eq:avdeltat}
\int_{\mathbb{T}^m}{\mathrm{d} \phi}\ \left({\Delta E}\left(v_0, b, M,\phi, c\right)+{%
\Delta E}\left(-v_0, b, M,\phi, c\right)\right)&=&  \notag \\
&\ &\hspace{-4.3cm}\frac{2\widehat{\beta^{\left(4\right)}_{II}}\left(e_0, b, M, c\right)}{\| v_0\|^4}%
+\mathrm{O}\left(\|v_0\|^{-5}\right).
\end{eqnarray}
Here
\begin{eqnarray}  \label{eq:beta4}
\widehat{\beta^{\left(4\right)}_{II}}\left(e_0,b, M,c\right)&=&\frac{c^2}{2}\int_{\mathbb{T}^m}{%
\mathrm{d} \phi}\int_0^1\mathrm{d}\lambda\int_0^1\mathrm{d}\lambda^{\prime}\
\left(\lambda-\lambda^{\prime}\right)^2\qquad \qquad \notag \\
&\ &\times\partial_\tau S\left(\lambda-\frac12,M,\phi\right)\partial_\tau
S\left(\lambda^{\prime}-\frac12,M,\phi\right),
\end{eqnarray}
with $S\left(\mu,M,\phi\right)=\nabla W\left(M^{-1}\left(b+\mu e_0\right), \phi\right)$.
\end{proposition}

We remark that $\Delta E\left(v_0, b, M, \phi, c\right)$ and $\Delta E\left(-v_0, b, M,\phi, c\right)$
are the energy changes undergone by two distinct particles, both impinging
at the same time on the same obstacle with the \emph{same} impact parameter,
but with opposite velocities. According
to Proposition~\ref{prop:deltat}, each of the two terms ${\Delta E}\left(\pm v_0,
b, M,\phi,c\right)$ is of order $\|v_0\|^{-1}$ so that Proposition~\ref{prop:avdeltat}
shows that combining a time average with a ``time reversal'' $v_0\to-v_0$
diminishes the energy change undergone by the particle in a scattering event
drastically.

\begin{proof}
Using (\ref{eq:deltakdeltaE}), we write, as in (\ref{eq:deltaenergyI})-(\ref{eq:deltaenergyII})
\begin{equation}\label{eq:splitenergy}
\Delta E=\Delta E_I +\Delta E_{II}+\mathrm{O}\left(\|v_0\|^{-5}\right).
\end{equation}
It is then immediately clear from (\ref{eq:deltaenergyI}) that
\begin{eqnarray}\label{eq:10}
\int_{\T^m}\rd\phi\ {\Delta E_{I}}\left(v_0, b, M,\phi, c\right)&=&\frac{c}{\|v_0\|}\int \rd\lambda\int\rd\phi\ \qquad\qquad\nonumber\\
&\ &\hspace{-4.5cm} \times\partial_\tau W\left(M^{-1}\left(b+\left(\lambda-\frac12\right) e_0\right), \omega\tau_0+\frac{\omega\lambda}{\|v_0\|}+\phi\right)=0,\nonumber
\end{eqnarray}
since $\partial_\tau =\omega\cdot\nabla_\phi$ and $W$ is $\phi$-periodic.

We now turn to ${\Delta E_{II}}$ and write (see (\ref{eq:deltaEexpansion}))
\begin{equation}\label{eq:betasplit}
\Delta E_{II}=\frac{\beta^{\left(3\right)}_{II}}{\|v_0\|^{3}}+\frac{\beta^{\left(4\right)}_{II}}{\|v_0\|^{4}}+\mathrm{O}\left(\|v_0\|^{-5}\right).
\end{equation}
One readily finds
\begin{eqnarray}\label{eq:3}
\beta^{\left(3\right)}_{II}\left(e_0, b, M,\phi, c\right)&=&-{c^2}\int_0^1\rd \lambda \int_0^\lambda \rd \lambda' \int_0^{\lambda'}
\rd \lambda''\qquad\qquad\qquad\nonumber\\
&\ &\hspace{-1cm}  \partial_\tau S\left(\lambda-\frac12,M,\phi\right)\cdot  S\left(\lambda''-\frac12, M,\phi\right).
\end{eqnarray}
and, immediately performing the $\phi$-average,
\begin{eqnarray*}
 \int_{\T^m}\rd \phi\ {\beta^{\left(4\right)}_{II}}\left(e,b, M,\phi, c\right)&=&\frac{c^2}{2}\int_{\T^m}\rd \phi\int_0^1\rd\lambda\int_0^1\rd\lambda'\  \left(\lambda-\lambda'\right)^2\qquad\qquad\nonumber\\
&\ &\quad\partial_\tau S\left(\lambda-\frac12,M,\phi\right)\partial_\tau S\left(\lambda'-\frac12,M,\phi\right).
\end{eqnarray*}
Note that in (\ref{eq:3}), the integrand is in general no longer a gradient in the $\phi$-variables, except in the special case where
$W\left(y,\phi\right)=w\left(y\right)f\left(\phi\right)$. So there is no reason why the $\phi$-average of $\beta^{\left(3\right)}_{II}\left(e_0, b, M,\phi, c\right)$ should vanish. But now remark, using (\ref{eq:3}) and the definition of $S$, that
\begin{eqnarray*}
\beta^{\left(3\right)}_{II}\left(-e_0, b, M,\phi, c\right)&=&-{c^2}\int_0^1\rd \mu\ \partial_\tau S\left(\frac12-\mu,M,\phi\right)\nonumber\\
&\ &\quad\cdot \int_0^\mu \rd \mu' \int_0^{\mu'}
\rd \mu''   S\left(\frac12-\mu'', M,\phi\right).\nonumber\\
\end{eqnarray*}
When performing in this last expression the succession of changes of variable defined by  $\frac12-\mu''=\lambda-\frac12$, $\mu'=1-\lambda'$, $-\mu+1=\lambda''$, one finds
\begin{eqnarray}\label{eq:4}
\beta^{\left(3\right)}_{II}\left(-e_0, b, M,\phi, c\right)&=&-{c^2}\int_0^1\rd \lambda''\ \int_{\lambda''}^{1} \rd \lambda' \int_{\lambda'}^{1}\rd\lambda \nonumber\\
&\ &\hspace{-3cm}\partial_\tau S\left(\lambda''-\frac12,M,\phi\right)\cdot   S\left(\lambda-\frac12, M,\phi\right).
\end{eqnarray}
Note that the domain of integration is the same as in (\ref{eq:3}), just the order of integration is different. So adding (\ref{eq:3}) and (\ref{eq:4}) the integrand becomes
$$
S\left(\lambda-\frac12, M,\phi \right)\partial_\tau S\left(\lambda''-\frac12, M,\phi\right)+ \partial_\tau S\left(\lambda-\frac12, M,\phi\right)S\left(\lambda''-\frac12, M,\phi\right)
$$
which is a total time derivative. Computing the $\phi$-average of the sum therefore yields
\begin{equation}\label{eq:11}
\int_{\T^m}\rd \phi\ \left({\beta^{\left(3\right)}_{II}}\left(e_0, b, M, \phi, c\right)+{\beta^{\left(3\right)}_{II}}\left(-e_0, b, M,\phi, c\right)\right)=0.
\end{equation}
A similar computation shows that
\begin{equation}\label{eq:timereversalok}
 \int_{\T^m}\rd \phi\ {\beta^{\left(4\right)}_{II}}\left(-e_0,b, M,\phi, c\right)=\int_{\T^m}\rd\phi\ {\beta^{\left(4\right)}_{II}}\left(e_0,b, M,\phi, c\right).
\end{equation}
Adding the various contributions, the proposition now follows from (\ref{eq:splitenergy}).
\end{proof}

We are now ready to prove Theorem~\ref{propo_2}.\newline
\noindent\textit{Proof of Theorem~\ref{propo_2}.} (i) Noting that
$\nabla\hat{W}\left(y,\tau\right)=M\nabla W\left(M^{-1}y,\omega \tau+\phi\right)$ one finds
\begin{equation*}
\int_{-\infty}^{+\infty} \rd \lambda \int_{e\cdot b=0}
\rd b\ \nabla\hat W\left(b+\left(\lambda-\frac12\right)e,\tau\right)=0
\end{equation*}
since $\hat W$ has compact support in its first variable. So $\overline{\alpha^{(1)}}=0$Similarly, integration
over the $\phi$ variable leads to the vanishing of $\overline{\alpha^{\left(2\right)}\left(e\right)}$.
To prove (\ref{eq:energyaverage}), we first point
out that, in view of the rotational invariance of the system, 
$
\Delta E\left(M^{\prime}v_0, M^{\prime}b, M^{\prime}M, \phi, c\right)=\Delta E\left(v_0, b,
M, \phi, c\right),
$
for
all $M^{\prime}\in\mathrm{SO}\left(d,\mathbb{R}\right)$.
Consequently, $\overline{\Delta E\left(M^{\prime}v_0\right)} =\overline{\Delta E\left(v_0\right)},$
where $\overline{\ \cdot\ }$ is defined in (\ref{eq:overlinedef}).
As a result,  $\overline{\Delta E\left(v_0\right)}$ depends only on $\|v_0\|$ and not
on $e_0$. In particular $\overline{\Delta E\left(-v_0\right)} =\overline{\Delta E\left(v_0\right)}.$
It therefore follows from (\ref{eq:avdeltat}) that
\begin{eqnarray*}
\overline{\Delta E\left(v_0\right)}=\frac{\overline{\beta^{\left(4\right)}_{II}}}{\|v_0\|^{4}}%
+\mathrm{O}\left(\|v_0\|^{-5}\right).
\end{eqnarray*}
This proves the first equation in (\ref{eq:energyaverage}). The second is
obtained similarly and equation (\ref{eq:betaaverage}) then follows
immediately. It remains to show (\ref{eq:bdlink}) and (\ref{eq:d2formula}).
For that purpose, we compute $B$:
\begin{eqnarray*}
B&=&\overline{\beta_{II}^{\left(4\right)}} = \frac{\overline{c^2}}{2}\int_{\mathbb{T}^{m}}d\phi \int_{\mathbb{S}%
^{d}}d\Omega\left(e_0\right)\int_{b\cdot e_0=0}\frac{db}{C_d}\int_0^1d
\lambda\int_0^1d\lambda^{\prime}\left(\lambda-\lambda^{\prime}\right)^2 \\
&\ &\qquad\qquad\qquad\qquad \tilde{S}\left(b+\left(\lambda-\frac{1}{2}\right)e_0,\phi\right)\cdot
\tilde{S}\left(b+\left(\lambda^{\prime}-\frac{1}{2}\right)e_0,\phi\right)
\end{eqnarray*}
where $\tilde{S}\left(y,\phi\right)=\nabla\partial_\tau W\left(y,\phi\right).$
Using  (\ref{eq:changevar}) below, we get
\begin{equation*}
B=\frac{\overline{c^2}}{2C_d}\int_{\mathbb{T}^{m}}d\phi  \int_{\mathbb{R}%
^d}dy \int_{\mathbb{R}^d}dy^{\prime}\left\|y-y^{\prime}\right\|^{3-d}
\tilde{S}\left(y,\phi\right)\cdot \tilde{S}\left(y^{\prime},\phi\right),
\end{equation*}
from which, using the definition of $\tilde{S}$ and integrating by parts, we
conclude
\begin{equation*}
B=\left(d-3\right)\frac{\overline{c^2}}{2C_d}\int_{\mathbb{T}^{m}}d\phi  \int_{\mathbb{R%
}^d}dy \int_{\mathbb{R}^d}dy^{\prime}\left\|y-y^{\prime}\right\|^{1-d}
\partial_\tau W\left(y,\phi\right)\partial_\tau W\left(y^{\prime},\phi\right).
\end{equation*}
Remarking now that
\begin{eqnarray*}
\left(\Delta E\left(v_0,\kappa\right)\right)^2
&=& \frac{c^2}{\left\|v_0\right\|^2}\int_{\mathbb{R}%
}d\lambda\int_{\mathbb{R}}d\lambda^{\prime}\ \partial_\tau
W\left(M^{-1}\left(b+\left(\lambda-\frac{1}{2}\right)e_0\right),\phi\right) \\
&\ &\quad\times\partial_\tau W\left(M^{-1}\left(b+\left(\lambda^{\prime}-%
\frac{1}{2}\right)e_0\right),\phi\right) + \mathrm{O}\left(\|v_0\|^{-3}\right)
\end{eqnarray*}
and using again (\ref{eq:changevar}) we finally find (\ref{eq:bdlink}) and (%
\ref{eq:d2formula})

(ii) Note that when $v_n$ is defined
by (\ref{eq:velrw}), $\kappa_n$ is independent of $v_n$ since
the latter only depends
on $\kappa_k$ for $k<n$.
It follows therefore from (\ref{eq:betaaverage}) that
$
\left\langle \beta_n^{\left(\ell\right)}\right\rangle=0=\left\langle \beta_n^{\left(4\right)}\right\rangle-B.
$
The same remark applies to the computation of the correlations. For example,
when computing
$
\left\langle \beta^{\left(\ell\right)}_n\beta^{\left(\ell^{\prime}\right)}_{n+k}\right\rangle,
$
for some positive $k$, one can integrate first over $\kappa_{n+k}$,
which yields the result because of (i). \qed

\begin{lemma}
In dimension $d\geq 2,$ for all $f:\mathbb{R}^d\times\mathbb{R}^d\times%
\mathbb{R}^+\rightarrow\mathbb{R},$ such that $\| y_0-y_0^{\prime}\|^{1-d}
f\left(y_0,y^{\prime}_0,\left\|y_0-y^{\prime}_0\right\|\right)\in L^1\left(\mathbb{R}^{2d}\right)$%
, we have
\begin{eqnarray}
&\int_{\mathbb{S}^d}d\Omega\left(e_0\right)\int_{b\cdot e_0=0}\mathrm{d} b\int_{\mathbb{%
R}}\mathrm{d}\lambda\int_{\mathbb{R}}\mathrm{d}\lambda^{\prime}f\left(b+\lambda
e_0,b+\lambda^{\prime}e_0,\left|\lambda-\lambda^{\prime}\right|\right)  \notag \\
&\qquad \qquad =\int_{\mathbb{R}^d}dy_0\int_{\mathbb{R}^d}dy^{\prime}_0\left%
\|y_0-y^{\prime}_0\right\|^{1-d}f\left(y_0,y^{\prime}_0,\left\|y_0-y^{\prime}_0%
\right\|\right).  \label{eq:changevar}
\end{eqnarray}
\end{lemma}

\begin{proof}

Let $y_0$ and $y_0'$ be in $\mathbb{R}^d,$ $y_0\not= y_0'$. Then there exists a unique
$\left(\lambda,\lambda',e_0,b\right)\in\mathbb R\times\mathbb
R\times\mathbb S^{d}\times \mathbb R^d$ with $e_0\cdot e_0=1$ such that
$$y_0=b+\lambda
e_0,\ y'_0=b+\lambda' e_0\ \textrm{and}\ b\cdot e_0=0.
$$
Since $e_0\in\mathbb S^{d},$ there exists unique angles
$\left(\theta_1,\cdots,\theta_{d-1}\right)\in[0,\pi]^{d-2}\times[0,2\pi]$ such
that
$$e_0=Ru_1,\ R=R_{d-1}\left(\theta_{d-1}\right)\cdots R_{1}\left(\theta_{1}\right),$$
where $\left(u_1,\ldots,u_d\right)$ is the canonical basis of $\mathbb{R}^d$ and
$R_i\left(\theta\right)$ is the rotation of angle $\theta$ in the plan defined by
$u_i$ and $u_{i+1}.$
Since $R^{-1}b$ is orthogonal to $u_1,$ there exists also a unique
$\left(\rho,\tilde\theta_2,\cdots,\tilde\theta_{d-1}\right)\in\mathbb{R}^+\times[0,\pi]^{d-3}\times[0,2\pi]$
such that
$$
b=\rho R\tilde R u_2,\ \tilde R=R_{d-1}\left(\tilde\theta_{d-1}\right)\cdots R_{2}\left(\tilde\theta_{2}\right).
$$
This gives the following equality:
$$
\rd y_0 \rd y'_0=\left|J\right| \rd\lambda \rd\lambda'\rd\rho \prod_{i=1}^{d-1}d\theta_i
\prod_{j=2}^{d-1}d\tilde\theta_j,
$$
where
$$
\left|J\right|=\left|\begin{tabular}{ccccc}
  $R u_1$ & $0_{d\times 1}$ & $R\tilde R u_2$ & $N$ & $M$ \\
  $0_{d\times 1}$ & $R u_1$ & $R\tilde R u_2$ & $N$ & $M'$
\end{tabular}
\right|,$$
with $N=\rho R\nabla_{\tilde \theta}\left(\tilde R u_2\right)$,
$$
M=\nabla_\theta\left(R\left(\rho\tilde R u_2+\lambda u_1\right)\right) \
\textrm{and}\ M'=\nabla_\theta\left(R\left(\rho\tilde R u_2+\lambda'
  u_1\right)\right).$$
Simple manipulations on the rows and columns yield
$$
\left|J\right|=\left|\begin{tabular}{ccc|cc}
  $R u_1$ & $R\tilde R u_2$ & $N$ & $0_{d\times 1}$ & $M$ \\

\hline

  \multicolumn{3}{c|}{$0_{d\times d}$}& $R u_1$  & $M'-M$
\end{tabular}
\right|=\left|\lambda'-\lambda\right|^{d-1}\rho^{d-2}J_1 J_2,$$
with
$
J_1=\left|Ru_1; \nabla_\theta R u_1\right|,\ J_2=\left|u_1;\tilde R u_2; \nabla_{\tilde \theta}\tilde R
  u_2\right|,
$
the result follows upon noticing that
$$
\rd\Omega\left(e_0\right)=J_1\prod_{i=1}^{d-1}\rd\theta_i\ \mathrm{and}\
\rd b=\rho^{d-2}J_2d\rho\prod_{j=2}^{d-1}\rd \tilde\theta_j.
$$

\end{proof}

\noindent\textit{Proof of Theorem~\ref{propo_3}.}
Computing $R\left(v, \kappa\right)$ to second order in perturbation theory, one finds
\begin{equation}\label{eq:Rvexpansion}
 R\left(v,\kappa\right)=R_I\left(v,\kappa\right)+R_{II}\left(v,\kappa\right),
\end{equation}
where
\begin{equation*}
 R_I\left(v,\kappa\right)=\frac{c}{\|v\|}\int\rd \mu\
 \hat g\left(b+\mu e, \tau_0 + \frac{\mu+\frac12}{\|v\|}\right)\quad \mathrm{and}
\end{equation*}
\begin{equation*}
 R_{II}\left(v,\kappa\right)=\frac{c^2}{\|v\|^3}\int \rd \mu\ \left[K\left(e,\kappa, \mu\right)\cdot \nabla\right] \hat g\left(b+\mu e, \tau_0\right)+\mathrm{O}\left(\|v\|^{-4}\right)
\end{equation*}
with
\begin{equation*}
 K\left(e,\kappa,\mu\right)=\int_{-\infty}^\mu \rd \mu' \int_{-\infty}^{\mu'} \rd \mu'' \hat g\left(b+\mu'' e, \tau_0\right).
\end{equation*}
As a result of (\ref{eq:centeredc}), $\overline{R_I\left(v,\kappa\right)}=0$, immediately implying $\overline{\alpha^{\left(\ell\right)}}=0$ for $\ell=1,2$ and hence $\overline{\beta^{\left(\ell\right)}}=0$ (see (\ref{eq:betaell})$\ell=0,1$. To compute $\overline{\beta^{\left(2\right)}}$, we need $\overline{e\cdot \alpha^{\left(3\right)}}$. From (\ref{eq:Rvexpansion}) we find
$
 \overline{e\cdot \alpha^{\left(3\right)}\left(e,\kappa\right)}=\overline{T\left(e,\kappa\right)}
$
with
\begin{eqnarray*}
 T\left(e,\kappa\right)&=&c^2\int_{-1}^1\rd\mu\ \int_{-1}^\mu\rd \mu'\ \int_{-1}^{\mu'}\rd\mu''\ \left[\hat g\left(b+\mu''e, \tau_0\right)\cdot\nabla\right]\left(e\cdot \hat g\right)\left(b+\mu e, \tau_0\right)\\
&=&c^2\int_{-1}^1\rd\mu\  \int_{-1}^{\mu}\rd\mu''\ \left(\mu-\mu''\right)\left[\hat g\left(b+\mu''e, \tau_0\right)\cdot\nabla\right]\left(e\cdot \hat g\right)\left(b+\mu e, \tau_0\right).
\end{eqnarray*}
Noting that the integrand is unchanged under the change of variable $\tilde e=-e, \tilde \mu =-\mu, \tilde \mu''=-\mu''$, one finds
\begin{eqnarray*}
 \int \rd\Omega\left(e\right)\int_{b\cdot e=0} \rd b\ T\left(e,\kappa\right)&=&c^2\int \rd\Omega\left(\tilde e\right)\int_{b\cdot \tilde e=0}\rd b\ \int_{-1}^1\rd\tilde\mu\ \int_{\tilde\mu}^1\rd\tilde\mu''\ \left(\tilde\mu-\tilde\mu''\right)\\
&\ &\quad\quad\quad \times\left[\hat{g}\left(b+\tilde\mu''\tilde e, \tau_0\right)\cdot \nabla\right]\left(\tilde e\cdot \hat g\right)\left(b+\tilde \mu \tilde e, \tau_0\right).
\end{eqnarray*}
Adding the last two formulas and using the change of variables formula (\ref{eq:changevar}), we conclude
\begin{equation*}
\overline{e\cdot \alpha^{\left(3\right)}}= \frac{\overline{c^2}}{2C_d}\int_{\T^m}{\rd\phi}\int \rd y\rd y''
\|y-y''\|^{1-d}\sum_j\left(y-y''\right)_j\left[g\left(y'',\phi\right)\cdot \nabla\right]g_j\left(y,\phi\right),
\end{equation*}
so that a partial integration yields
\begin{eqnarray*}
\overline{e\cdot \alpha^{\left(3\right)}}&=& -\frac{\overline{c^2}\left(1-d\right)}{2C_d}\int_{\T^m}{\rd\phi}\int \rd y\rd y''
\|y-y''\|^{-1-d}\\
&\ &\hspace{3cm}\times\left(\left(y-y''\right)\cdot g\left(y,\phi\right)\right)\left( \left(y-y''\right)\cdot g\left(y'',\phi\right)\right)\\
&\ &\qquad -\frac{\overline{c^2}}{2C_d}\int_{\T^m}{\rd\phi}\int \rd y\rd y''
\|y-y''\|^{1-d} g\left(y,\phi\right) \cdot g\left(y'',\phi\right).
\end{eqnarray*}
From (\ref{eq:alpha1}) one easily sees the second term equals $-\frac12\overline{\alpha^{\left(1\right)}\cdot\alpha^{\left(1\right)}\left(e\right)}$ so that we find, using (\ref{eq:betaell})
\begin{eqnarray*}
 \overline{\beta^{\left(2\right)}\left(e\right)} &=& \frac{\overline{c^2}\left(d-1\right)}{2C_d}\int_{\T^m}{\rd\phi}\int \rd y\rd y''
\|y-y''\|^{-1-d}\\
&\ &\hspace{3cm}\times\left(\left(y-y''\right)\cdot g\left(y,\phi\right)\right)\left( \left(y-y''\right)\cdot g\left(y'',\phi\right)\right)\\
&=&\frac{d-1}{2}\overline{\beta^{\left(0\right)}\left(e\right)^2}\geq 0.
\end{eqnarray*}
\qed


\bibliographystyle{alpha}
\bibliography{randombibl}

\begin{thebibliography}{AWMN06}

\bibitem[AWMN06]{awmn}
E.~Arvedson, M.~Wilkinson, B.~Mehlig, and K.~Nakamura.
\newblock Staggered ladder operators.
\newblock {\em Phys. Rev. Lett.}, 96:030601, 2006.

\bibitem[BMV08]{bmv}
A.~Beck and N.~Meyer-Vernet.
\newblock The trajectory of an electron in a plasma.
\newblock {\em Am. J. Phys.}, (10):934--936, 2008.

\bibitem[BSC90]{bcs}
L.~A. Bunimovich, Ya.~G. Sinai, and N.I. Chernov.
\newblock Statistical properties of two-dimensional hyperbolic billiards.
\newblock {\em Russ. Math. Surv.}, 45:105--152, 1990.

\bibitem[DK09]{dk}
D.~Dolgopyat and L.~Koralov.
\newblock Motion in a random force field.
\newblock {\em Nonlinearity}, (22):187--211, 2009.

\bibitem[Eij97]{ve}
E.~Vanden Eijnden.
\newblock Some remarks on the quasilinear treatment of the stochastic
  acceleration problem.
\newblock {\em Phys. Plasmas}, 4(5):1486--1488, 1997.

\bibitem[GFZ91]{gff}
L.~Golubovi\ifmmode\acute{c}\else\'{c}\fi{}, S.~Feng, and F.~Zeng.
\newblock Classical and quantum superdiffusion in a time-dependent random
  potential.
\newblock {\em Phys. Rev. Lett.}, 67(16):2115--2118, 1991.

\bibitem[GR08]{gr}
T.~Goudon and M.~Rousset.
\newblock Stochastic acceleration in an inhomogenous time random force field.
\newblock {\em preprint}, 2008.

\bibitem[Hei92]{h92}
J.~Heinrichs.
\newblock Diffusion and superdiffusion of a quantum particle in time-dependent
  random potentials.
\newblock {\em Z. Phys. B}, 89:115--121, 1992.

\bibitem[JK82]{jk}
A.~M. Jayannavar and N.~Kumar.
\newblock Nondiffusive quantum transport in a dynamically disordered medium.
\newblock {\em Phys. Rev. Lett.}, 48(8):553--556, 1982.

\bibitem[KP79]{kp79}
H.~Kesten and G.~C. Papanicolaou.
\newblock A limit theorem for turbulent diffusion.
\newblock {\em Comm. Math. Phys.}, 65(2):97--128, 1979.

\bibitem[LBP]{ldp}
P.~Lafitte, S.~De Bi\`evre, and P.~Parris.
\newblock Normal transport properties in a metastable stationary state for a
  classical particle coupled to a non-ohmic bath.
\newblock {\em J. Stat. Phys.}

\bibitem[LMF95]{lmf}
N.~Lebedev, P.~Maas, and S.~Feng.
\newblock Diffusion and superdiffusion of a particle in a random potential with
  finite correlation time.
\newblock {\em Phys. Rev. Lett.}, 74(11):1895--1899, 1995.

\bibitem[PV03]{pv}
Fr{\'e}d{\'e}ric Poupaud and Alexis Vasseur.
\newblock Classical and quantum transport in random media.
\newblock {\em J. Math. Pures Appl. (9)}, 82(6):711--748, 2003.

\bibitem[Ros92]{r}
M.N. Rosenbluth.
\newblock Comment on ``\protect{C}lassical and quantum superdiffusion in a
  time-dependent random potential.
\newblock {\em Phys. Rev. Lett.}, 69:1831, 1992.

\bibitem[RY99]{ry}
Daniel Revuz and Marc Yor.
\newblock {\em Continuous martingales and {B}rownian motion}, volume 293 of
  {\em Grundlehren der Mathematischen Wissenschaften [Fundamental Principles of
  Mathematical Sciences]}.
\newblock Springer-Verlag, Berlin, third edition, 1999.

\bibitem[SG73]{sg}
M.B. Silevitch and K.I. Golden.
\newblock Dielectric formulation of test particle energy loss in a plasma.
\newblock {\em J. Stat. Phys.}, 7(1):65--87, 1973.

\bibitem[SPB06]{spd}
A.A. Silvius, P.E. Parris, and S.~De Bi\`{e}vre.
\newblock Adiabatic-nonadiabatic transition in the diffusive hamiltonian
  dynamics of a classical holstein polaron.
\newblock {\em Phys. Rev. B}, 73:014304, 2006.

\bibitem[Stu65]{st}
P.~A. Sturrock.
\newblock Stochastic acceleration.
\newblock {\em Phys. Rev.}, 141(1):186--191, 1965.

\end{thebibliography}

\end{document}